%% file: M101pp.tex
\documentclass[12pt,preprint]{aastex}

\shorttitle{Luminous Stars in M101 }
\shortauthors{Grammer and Humphreys }

\begin{document}

\title{The Massive Star Population in M101. III. Spectra and Photometry of the Luminous and Variable Stars\altaffilmark{1}}

\author{Skyler H. Grammer\altaffilmark{2}, Roberta M.. Humphreys\altaffilmark{2} and, Jill Gerke\altaffilmark{3}}

\altaffiltext{1}{Based  on observations  with the Multiple Mirror Telescope, 
a joint facility of the Smithsonian Institution and the University of Arizona 
and on observations obtained with the Large Binocular Telescope (LBT), an international collaboration among institutions in the United
 States, Italy and Germany. LBT Corporation partners are: The University of
 Arizona on behalf of the Arizona university system; Istituto Nazionale di
 Astrofisica, Italy; LBT Beteiligungsgesellschaft, Germany, representing the
 Max-Planck Society, the Astrophysical Institute Potsdam, and Heidelberg
 University; The Ohio State University, and The Research Corporation, on
 behalf of The University of Notre Dame, University of Minnesota and
 University of Virginia.}

\altaffiltext{2}{Minnesota Institute for Astrophysics, 116 Church St SE, University of Minnesota , Minneapolis, MN 55455, grammer@astro.umn.edu, roberta@umn.edu} 

\altaffiltext{3}{Department of Astronomy, The Ohio State University, 140 West 18th Avenue, Columbus, OH 43210, USA}

\begin{abstract}
We discuss moderate resolution spectra, multicolor photometry, and light curves of
thirty-one of the most luminous stars and variables in the giant spiral M101. The 
majority are intermediate A to F-type supergiants. We present new photometry and 
light curves for three known ``irregular blue variables'' V2, V4 and V9)
and identify a new candidate. Their spectra and variability confirm that 
they are LBV candidates and V9 may be in
an LBV-like maximum light state or eruption. 
\end{abstract}

\keywords{galaxies: individual (M101) -- supergiants}

\section{Introduction} 

Recent supernova surveys have lead to the identification of an 
increasing number of non-terminal optical transients with a wide range 
of properties.  Some of these optical transients appear to be similar 
to the giant eruptions of the $\eta$ Car variables \citep{Humphreys:1999,Van-Dyk:2005,Van-Dyk:2012}, while others are more akin to the variability of normal Luminous Blue Variables (LBVs).  A very small fraction of the optical transients originate from lower luminosity, heavily obscured progenitors that may be extreme asymptotic giant branch (AGB) stars or in a post red supergiant stage of evolution \citep{Thompson:2009, Khan:2010, Bond:2011}.  The continued monitoring of these optical transients has led to the realization that in some cases the apparent terminal explosion is preceded by smaller eruptions, e.g. SN2005gl \citep{Gal-Yam:2007,Galyam09}, SN2006jc \citep{Pastorello:2007} and most recently the peculiar SN2009ip \citep{Mauerhan:2013, Pastorello:2013, Fraser:2013, Margutti:2013}.  Consequently, the connection between LBVs, giant eruptions, and true supernovae has come into question.  But very little is 
known about the origin of these giant eruptions, their progenitors and 
their evolutionary state.  An improved census of the most massive, 
evolved stars including the LBVs, and the hypergiants that occupy the upper HR Diagram is necessary to better characterize the properties of the 
possible progenitors. For these reasons,
we have begun a survey of the evolved massive star populations in 
several nearby galaxies  \citep{Humphreys:2013, Grammer:2013a, Humphreys:2014}.  

This paper is the third in a series on the massive star content of M101.  In the first two papers, we presented the photometric analysis and identification of the luminous and massive star populations, here we present spectroscopy and  multi-epoch imaging 
for the most luminous stars.  In the next section, we describe our target selection, observations, and data reduction.  In $\S3$ we discuss the stars for which we have spectra and light curves, and   in $\S4$, we present  the light curves for those without spectra.  We 
summarize our conclusions in the last section.

\section{Data and Observations}

Our motivation for this study is to examine the spectra and photometric variability
of the most luminous stars in M101.  Using their spectra and light curves,
we identify LBV candidates, hypergiants, and other luminous stars and emission-line stars.

\subsection{Target Selection}

Most of our targets were selected from the Hubble Legacy Archive (HLA) aperture photometry of \textit{HST}/ACS images from proposals GO-9490 (Nov. 2002) and GO-9492 (Jan. 2003) to be   brighter than $V\approx20.5$ mag.  Since the images of M101 are crowded, particularly in the spiral arms, we later performed our own photometry using Dolphot \citep{Dolphin:2000} to create a catalog of high precision photometry even in crowded regions \citep[hereafter Paper I]{Grammer:2013a}. In this paper,  photometry from Paper I is referred to as the catalog photometry.  We cross-identified targets selected from the HLA with the catalog using a tolerance of 0.1$\arcsec$ in radial separation.  We note that a few of the targets with $V < 20.5$ in the HLA photometry are much fainter in the catalog.  The targets with differences in $V$ larger than a few tenths of a magnitude are in regions where aperture photometry is inappropriate (e.g. crowded regions).  We visually inspected the surrounding region of each unmatched star and found in all cases, that  the unmatched targets were located in parts of the galaxy where photometry was likely to be highly compromised.

In addition to the targets selected from \textit{HST}/ACS photometry, we included blue supergiants and known luminous variables from \cite{Sandage:1974c} and \cite{Sandage:1983}.  The blue supergiants and luminous variables were originally identified on photographic images.  Since precise astrometry is required for our study, we used a Sloan Digital Sky Survey (SDSS) $g$ image of M101 to identify the blue supergiants and known variables by eye.  Many of the \cite{Sandage:1983} stars were also in regions of significant crowding which made positive identification difficult. Thus, we were only able to include 7 stars: B4, B53, B65, B162, V2, V4, and V9.

\subsection{Spectroscopy}

\begin{figure}
\figurenum{1} 
\epsscale{0.8}
\plotone{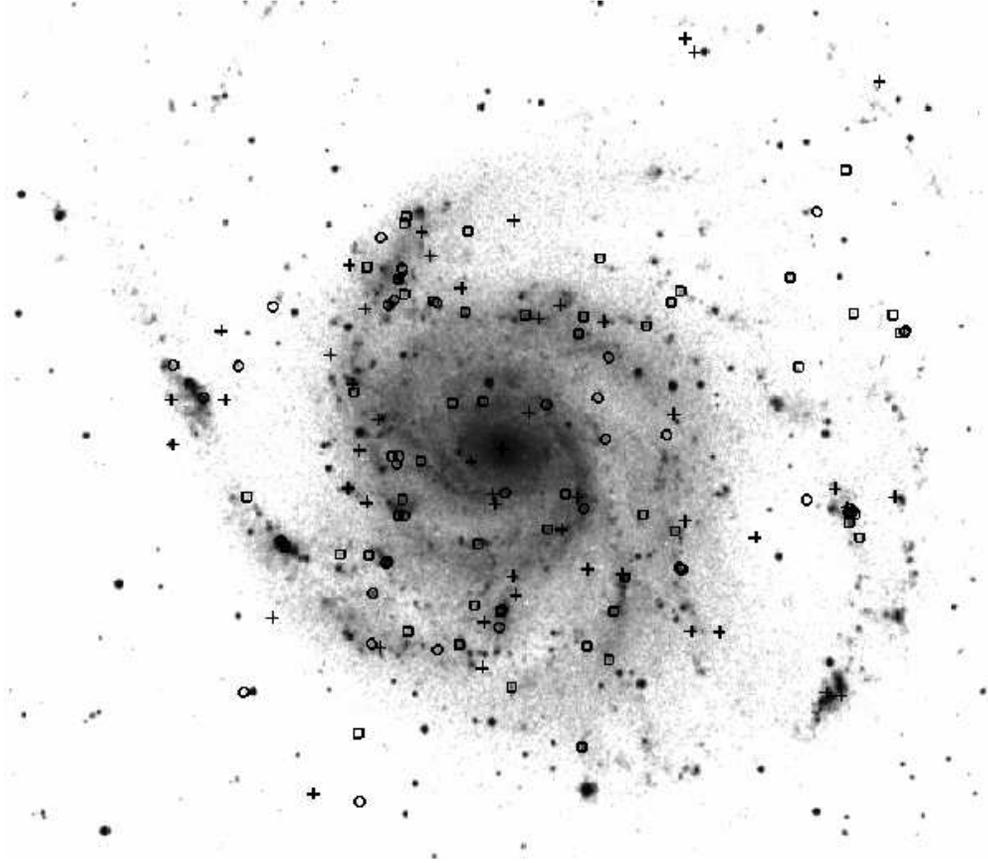}
\caption[POSS II red image of M101]{POSS II red image of M101.  Spectroscopic ta
rgets observed with MMT/Hectospec are shown as crosses and LBT/MODS as boxes.  T
argets without spectra are shown as circles.}
\label{figFOV}
\end{figure}

When we were selecting targets for spectroscopy, we did not yet have
the  light curves from  the Large Binocular Telescope (LBT) survey, 
discussed below \S {2.3}.   
 Since we could not obtain spectra for every star, we prioritized 
our  targets for spectroscopy by roughly estimating their variability using an ``absdiff'' image. 
  The absdiff image was created by taking the absolute value of the difference between a reference image, described below, and all other images.  Then, the subtracted images were convolved with a 2 pixel Guassian filter and summed.  In the absdiff image, the ``brighter'' the source, the more variable it is likely to be.  All stars with HLA $V$-band magnitudes 
  brighter than 20.5 mag were then overlaid onto the absdiff image.  
  Priority for spectroscopy was assigned by degree of variability 
  and $V$-band magnitude.  Thus the brightest stars with clear indications of variability received the highest priority.  We selected 56 of the brightest and most 
  variable stars for spectroscopy with the Hectospec on the Multiple Mirror Telescope (MMT), and 46 of the fainter ones for spectroscopy with the LBT MODS1 spectrograph. The spectrscopic targets are shown in Figure~\ref{figFOV}.

The MMT observations were obtained in June 2012 with the Hectospec multi-object spectrometer \citep[MOS;][]{Fabricant:1998}.  The 
Hectospec\footnotemark[1] is a fiber-fed MOS with a 1$^{\circ}$ FOV 
and 300 fibers; each fiber subtends $1.5\arcsec$ 
on the sky.  We used the 600 mm$^{-1}$ grating with the blue tilt centered on 4800{\AA} and the red tilt centered on 7300{\AA}.  
 The 600 mm$^{-1}$ grating gives  a spectral coverage of $\sim2500${\AA} 
 with 0.54{\AA} pixel$^{-1}$ resolution. The total integrated exposure times were 240 
 minutes in the blue and 180 minutes in 
the red.  The spectra were reduced using an exported version of the CfA/SAO SPECROAD package for Hectospec data E-SPECROAD\footnotemark[2].  The
specta were bias subtracted, flat -fielded, wavelength calibrated, 
and sky subtracted. The IRAF task \texttt{sensfunc}, in the \textit{ONEDSPEC} package, was used to flux calibrate the spectra with the standard Feige-66.  

\footnotetext[1]{http://www.cfa.harvard.edu/mmti/hectospec.html}
\footnotetext[2]{External SPECROAD was developed by Juan Cabanela for use on Linux or MacOS X systems outside of CfA. It is available online at http://iparrizar.mnstate.edu.}

The 46 stars selected for observation with the LBT were observed 
in May 2012 and June 2013 using the Multiple Object Dual Spectrograph (MODS) \citep{Pogge:2006}. For multi-object spectroscopy MODS uses masks with a 6.5 arcmin FOV, and  a dichroic with two gratings, 400 mm$^{-1}$ and 670 mm$^{-1}$, to get a full spectral range of 3200{\AA} to 10000{\AA}, and 
moderate resolution (R $\sim$ 1800).  Since the MODS FOV is considerably 
smaller than Hectospec, four masks were required to cover the disk of 
M101.  Problems with a beta version of the  reduction pipeline prevented us from  including 
the LBT spectra here.  
Analysis of the LBT/MODS spectra will be included in a future paper.  

The 31 confirmed members  are listed in order of right ascension in Table~\ref{tab:spec_members} with object identification, position, photometry, variability 
and their spectral type. The 19 foreground stars are 
in Table 2. The S/N in the spectra for six of the targets 
was too poor to assign a spectral type. The blue and red spectra of all of 
the targets are 
available at http://etacar.umn.edu/LuminousStars/M101/ in FITS format.
The flux calibrated and smoothed spectra are also available in a subdirectory.
Spectra of selected members are discussed in the next section.

\subsection{LBT Imaging}
M101 has been monitored as part of a variability survey of 27 nearby ($<10$ Mpc) galaxies using the twin 8.4m  LBT \citep{Kochanek:2008,Gerke}.  Between March 2008 and January 2013, M101 was observed using the Large Binocular Camera (LBC) in the $R$-band with the red-optimized LBC-Red camera while simultaneously cycling through observations in the $UBV$ filters with the blue-optimized LBC-Blue camera \citep{Giallongo:2008}.  One of us, J. Gerke,  did the basic data reduction steps including overscan correction, bias subtraction, and flat fielding with the IRAF \textit{MSCRED} package, as well as the subsequent analysis of the images.  Although M101 was observed even in sub-optimal conditions, only images with a point spread function (PSF) with a full width half max (FWHM) $\lesssim2\arcsec$ are analyzed.   

For the variability analysis, the ISIS image subtraction package \citep{Lupton:1998, Alard:2000} was used to process the multiple LBT images.  The image subtraction works by matching a reference image to each exposure and subtracting it to leave only the sources which have time-variable flux.  The reference image was created by median combining the four best seeing  epochs where an epoch is defined as the images for one night.  The astrometric solutions were determined using the IRAF package \textit{MSCTPEAK} and SDSS stars \citep{Ivezic:2007} in the FOV.  For each epoch, the $R$-band image serves as the astrometric reference image for all four filters which ensures identical astrometric solutions between filters.  By doing this, any ambiguity associated with cross-matching sources between filters is minimized.  The typical astrometric errors are $0.1\arcsec$.  

\begin{figure}
\figurenum{2}
\epsscale{0.8}
\plotone{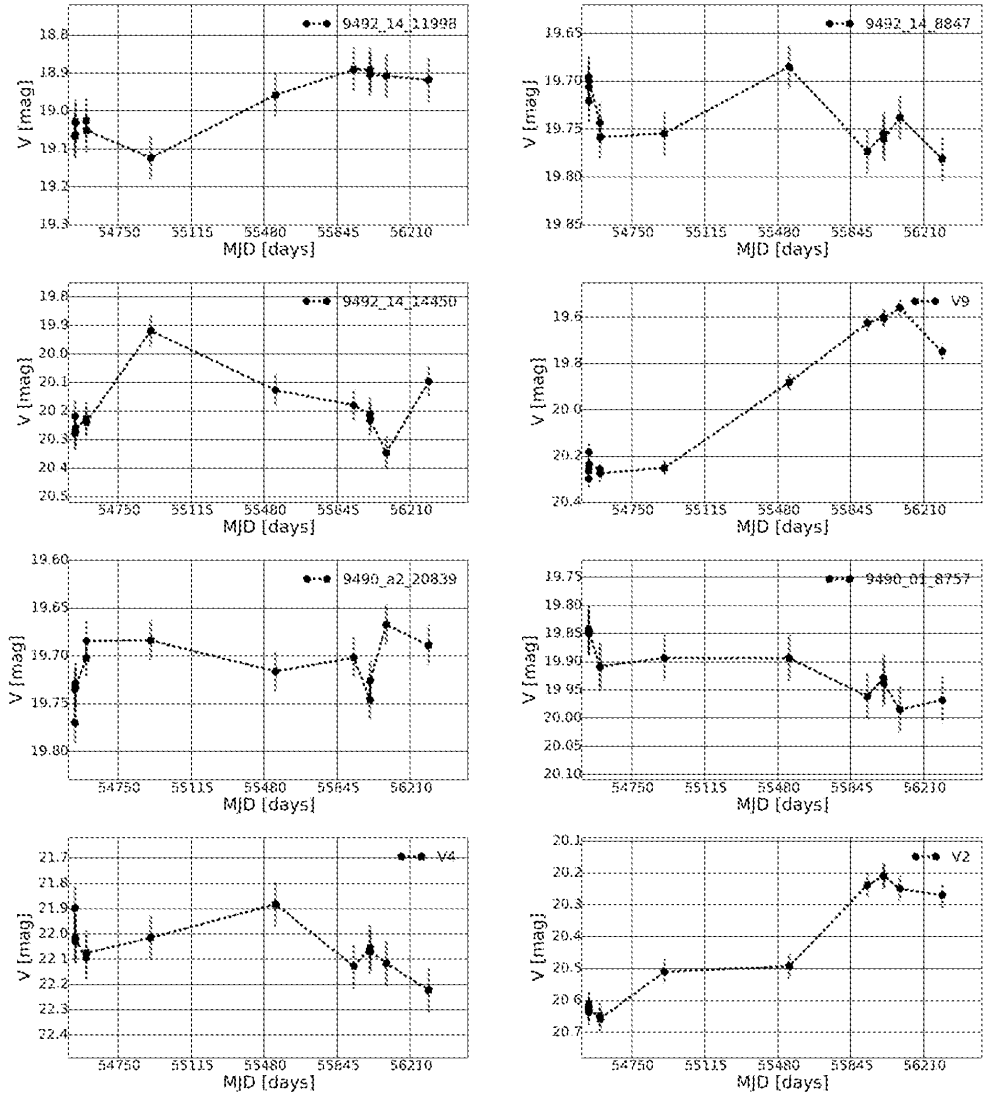} 
\caption{The LBT light curves for our stars with spectra  that also meet our cri
terion for variability.  Foreground stars are not shown
.  The x-axis grid lines correspond to 365 days.}
\label{figSpecLC}
\end{figure}

The light curves for the stars with HLA $V$-band magnitudes brighter than 20.5 mag  were extracted using ISIS.  Instrumental magnitudes for each epoch were converted to $UBVR$ magnitudes using photometric calibrations based on SDSS photometry which was transformed from the SDSS $ugriz$ filter system to the $UBVR$ system using the prescription described by \citet{Jordi:2006}.  The resulting photometry has photometric errors that are $\lesssim0.06$ magnitudes.  For each target, we calculate the root-mean-square (RMS) error, with respect to the median magnitude, as a measure of stellar variability.  We identify targets as variable sources for further analysis if their RMS variability is greater than the median photometric error.  The $V$-band light curves for spectroscopic targets observed with MMT which met our criterion for variability are shown in Figure~\ref{figSpecLC}

\section{Classification of the Stars}

For discussion, we have  grouped the spectra of the confirmed 
members by their broad spectral characteristics and known variability. 
In this section we describe the characteristics of four groups: the early-type 
or hot supergiants, intermediate-type  supergiants, emission-line sources, and candidate LBVs.  We describe the criteria for membership in each group in the  subsections below  and 
discuss the spectra, photometry, and light curves of several interesting 
or representative stars.  Group membership and  comments on their spectra and variability are also included in   Table~\ref{tab:spec_members}. 
For reference, we assume the  distance modulus, derived from the tip of the red giant branch, of (m-M)$_{0}$ = 29.05 $\pm$ 0.06(random) $\pm$ 0.12 (systematic) magnitudes \citep{Shappee:2011}. The foreground extinction towards M101 is only E$(B-V)$ = 0.01 \citep{Schlegel:1998} which corresponds to a visual extinction of $A_{V}$ = 0.03 magnitudes assuming a Galactic extinction law \citep{Cardelli:1989}.  For those stars with 
spectral types, we estimate the internal extinction by comparing their observed colors with  the intrinsic colors from \citet[see Table~\ref{tab:spec_members}]{Flower:1977,Flower:1996}.
Representative blue spectra from $\approx$3900 -- 5200{\AA}  are shown in Figures 3 to 6, and snapshot images of the confirmed members are in the Appendix.

\subsection{The Hot Supergiants}

This group includes the  luminous O- and B-type supergiants.  
Many of these stars show strong emission lines, mostly nebular, but with an
absorption line  spectrum  strong enough to allow  an estimate of the 
spectral type.  In this section we describe stars with interesting spectral features and/or photometric variability.

\begin{figure}
\figurenum{3}
\epsscale{1.0}
\plotone{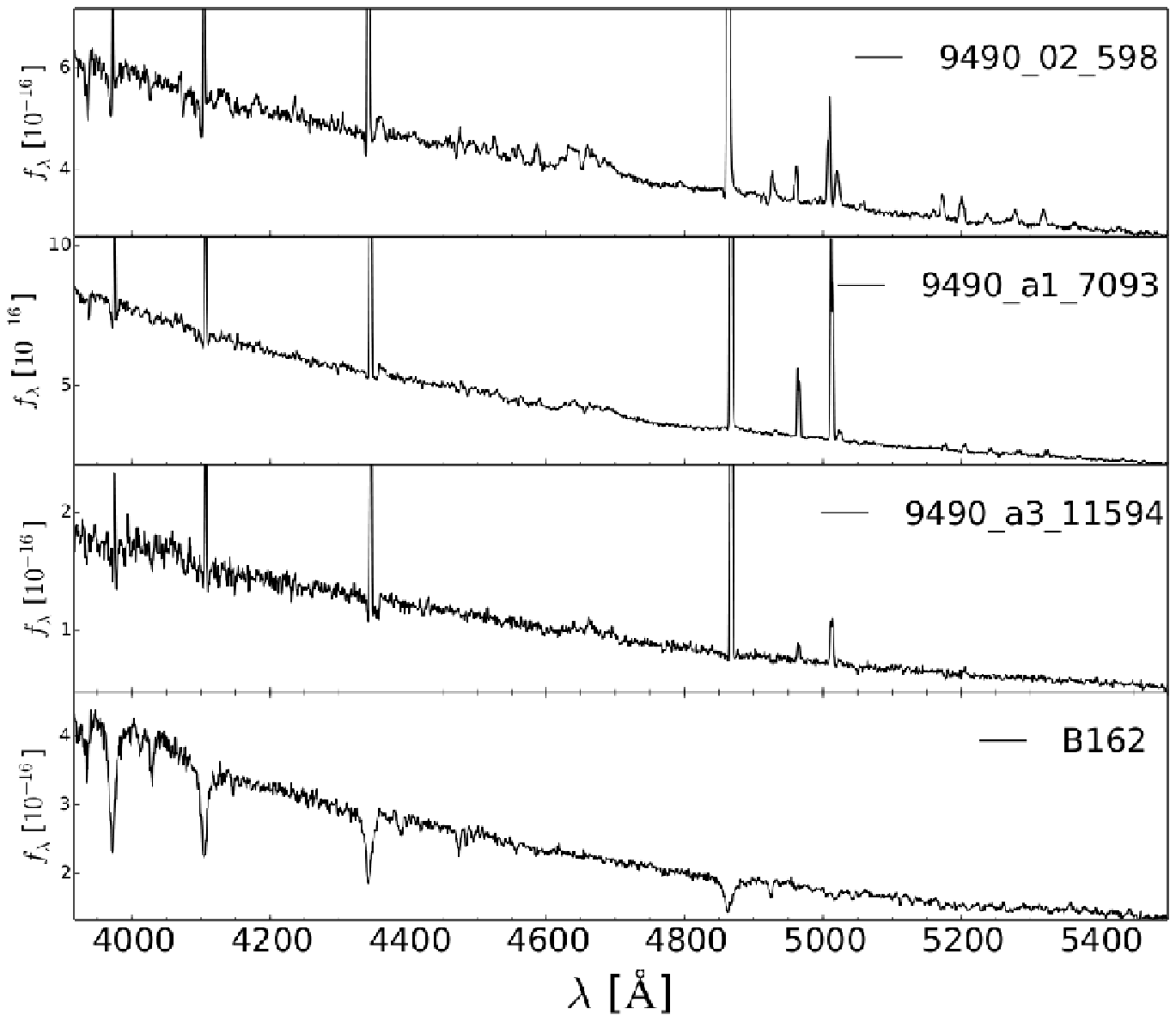}
\caption{Spectra of selected early-type supergiants. The spectra are flux calibr
ated and smoothed with a 3 box smooth in the {\it splot} task in IRAF.}
\label{figHotSp}
\end{figure}

The spectrum  of \textit{9490$\_$02$\_$598 (early B: I + WN + ?)} shows absorption lines of  Si IV $\lambda\lambda$4090 and 4116, Si III $\lambda$4552, C III $\lambda$4650, and He I $\lambda\lambda$4026 and 4144  consistent with an early B-type  supergiant (Figure~\ref{figHotSp}), although a Ca II K line is also present.  The Balmer series, as well as the He I lines of $\lambda\lambda$ 4471, 4922, 6678 and 7065  are in emission with P-Cygni profiles. He I $\lambda$ 5876 has a double or split emission profile not present in  any of the other emission lines.   Emission lines of Fe II, [Fe II], and the broad WN nitrogen emission features at $\lambda$4630-4670{\AA} and $\lambda$5680-5730{\AA} are present.  The terminal velocity, determined from P-Cygni profiles, is normally measured from the blue edge of the absorption component.  Because the spectra are moderate resolution with low S/N, we can more reliably determine the wind velocity from the absorption minima, see \citep{Humphreys:2014}.  We find a wind velocity of $392.0\pm12.6$ kms$^{-1}$ which is on the low end, but in the range for normal early-type supergiants \citep{Crowther:2006, Mokiem:2007}.

The catalog photometry for 9490$\_$02$\_$598 is $V$ = 20.07 and $(B-V)$ = 0.30.  Based on its early B-type spectrum , we assume an intrinsic  $(B-V)_{0}$ color of $\approx$ -0.20 \citep{Flower:1977, Flower:1996} yielding  a fairly high interstellar extinction of A$_{v}$ of 1.6 mag and M$_{v}$ = -10.5 mag.    Inspection of the \textit{HST}/ACS $V$ band image shows that 9490$\_$02$\_$598 may be a point source although it is located in a crowded region (Figure A10).  It  was not identified as a point source in the LBT images due to poorer spatial resolution, thus we do not have variability information.  The spectrum is  similar to an LBV in quiescence \citep{Humphreys:1994} but without a light curve we cannot say with any certainty that 9490$\_$02$\_$598 is an LBV candidate. With the Ca II K line it may be composite.    

The spectrum  of \textit{9490$\_$a3$\_$11594(Early B: I + WN:)} is shown  
in Figure~\ref{figHotSp}.  The presence of the C III/O II features at $\lambda\lambda$4068-4076, Si IV at $\lambda$4089 and $\lambda$4116 and absorption lines of He I at  $\lambda\lambda$4026, and 4144 indicate that it  has an early B-type spectrum. It also has a weak WN feature at $\lambda\lambda$ 4630-4670. 
The relative strength of the O II $\lambda\lambda$4070-4076 to He I $\lambda$4026 indicates high luminosity.  The emission lines are primarily nebular. We note that H$\beta$, He I $\lambda$5876 and the [O III] lines are double, but the [N II], [S II] and H$\alpha$ in the red are not. These double emission lines are discussed below. 

The catalog photometry for 9490$\_$a3$\_$11594 is $V$ = 21.5 and $(B-V)$ = 0.06, but  
the HLA and LBT photometry show it  $\approx$2 magnitudes brighter and $\approx$0.3 magnitudes redder than the catalog photometry.  9490$\_$a3$\_$11594 is located in a crowded region (Figure A12), thus  the  magnitudes obtained 
from aperture photometry may be significantly altered  by the neighboring stars.  The 
multi-epoch photometry for 9490$\_$a3$\_$11594 shows no variability in magnitude or color.

\textit{9490$\_$a1$\_$7093 (Mid B: I)}  has strong Balmer and 
nebular emission lines (Figure~\ref{figHotSp}). It  resembles 9490$\_$02$\_$598 with He I absorption lines $\lambda$4026, $\lambda$4144, and $\lambda$4387 and the C III/O II absorption feature. These lines suggest an early B-type spectrum,  but the  Mg II $\lambda$4481 seems too strong. It is in a crowded field so the spectrum may be a blend. We tentatively classify  it as a  mid-B-type supergiant.  Fe II and [Fe II] emission lines are present, as well as 
the WN nitrogen emission at $\lambda$ 4630-4670{\AA}.  He I $\lambda$ 5876{\AA}, H$\beta$, the [O III] nebular lines and the Fe II emission lines have double or split emission profiles. However, the nebular [N II] and [S II] lines in the red plus H$\alpha$ do not show the 
split profiles. 

The catalog color and magnitude for 9490$\_$a1$\_$7093 are $(B-V)$ = 0.31 and $V$ = 20.20, respectively.  9490$\_$a1$\_$7093 is located in a very crowded region (Figure A28) which the lower resolution of the LBT cannot resolve.  Consequently, we do not have information on its variability.  Since we were unable to assign a precise spectral type to 9490$\_$a1$\_$7093, we estimate the likely extinction for a mid-B type supergiant,
A$_{V}$ = 1.3 mag. This may seem high but 9490$\_$a1$\_$7093 is embedded in a region of intense star formation.   

\textit{B162 (B8: I)}   was identified as a blue supergiant in M101 by \cite{Sandage:1983}.  Followup spectroscopy by \cite{Humphreys:1987} confirmed its membership,
They estimated its  spectral type to be late B to early A, and suggested that B162 may be a composite based on the width of the Balmer lines.  Our spectrum of B162 (Figure~\ref{figHotSp}) is one of the few without any nebular contamination. Based on the   He I $\lambda$4471 and  Mg II $\lambda$4481 absorption lines, we suggest a spectral type of B8; the ratio of Si III $\lambda$4552 to He I $\lambda$4387, as well as the presence of the high luminosity indicator O I $\lambda$7774, confirms that B162 is a supergiant.

The catalog photometry for B162 indicates it is extremely luminous with a
$V$-band magnitude of $V$ = 19.52.  As a result, B162 is saturated in the LBT images thus preventing us from assessing its variability.  \cite{Sandage:1983} found no evidence for variability. The observed color for B162 is $(B-V)$ = 0.09 which is only 0.1 magnitudes redder than expected for its spectral type.  Correcting for extinction, its M$_{v}$ is -9.9 mag and with the bolometric correction for 
a late B-type  supergiant, we estimate the bolometric magnitude for B162 to be $M_{bol}$ = -10.4. The Balmer lines seem  too strong however for such a high luminosity star.  It may be a composite or blend of more than one star. See Figure A18, where the object appears to be an unresolved small group of 2 - 3 stars.

Three of the stars described in this subsection have double emission profiles in the nebular and hydrogen emission lines. We find that this is the case for several stars in this
study. Since the stars are in crowded regions with strong
nebular emission, the double features may be due to emission from the two sides of these large H II regions or from more than one emission region along the line of sight. 
However in the case of 9490$\_$a1$\_$7093 described above, the Fe II emission lines which presumably arise in the  stellar wind, also show the split profiles. For this star the velocity separation of the Fe II peaks is 180 km s$^{-1}$ which corresponds to a 
possible velocity of expansion of $\approx$ 90 km s$^{-1}$.

The spectral types, catalog photometry, visual extinction, and other notes  regarding 
the remaining hot or early-type supergiants are given in Table~\ref{tab:spec_members}.  

\subsection{The Intermediate-Type  Supergiants}
The intermediate supergiant group includes the visually most luminous A- to F-type 
stars.  The intermediate or yellow supergiants, often have strong emission due to stellar winds and mass loss,  but due to the degree of nebular contamination in most of these spectra, a stellar origin is uncertain in most cases.   Here we discuss  some representative and interesting stars in this group.

\begin{figure}
\figurenum{4}
\epsscale{1.0}
\plotone{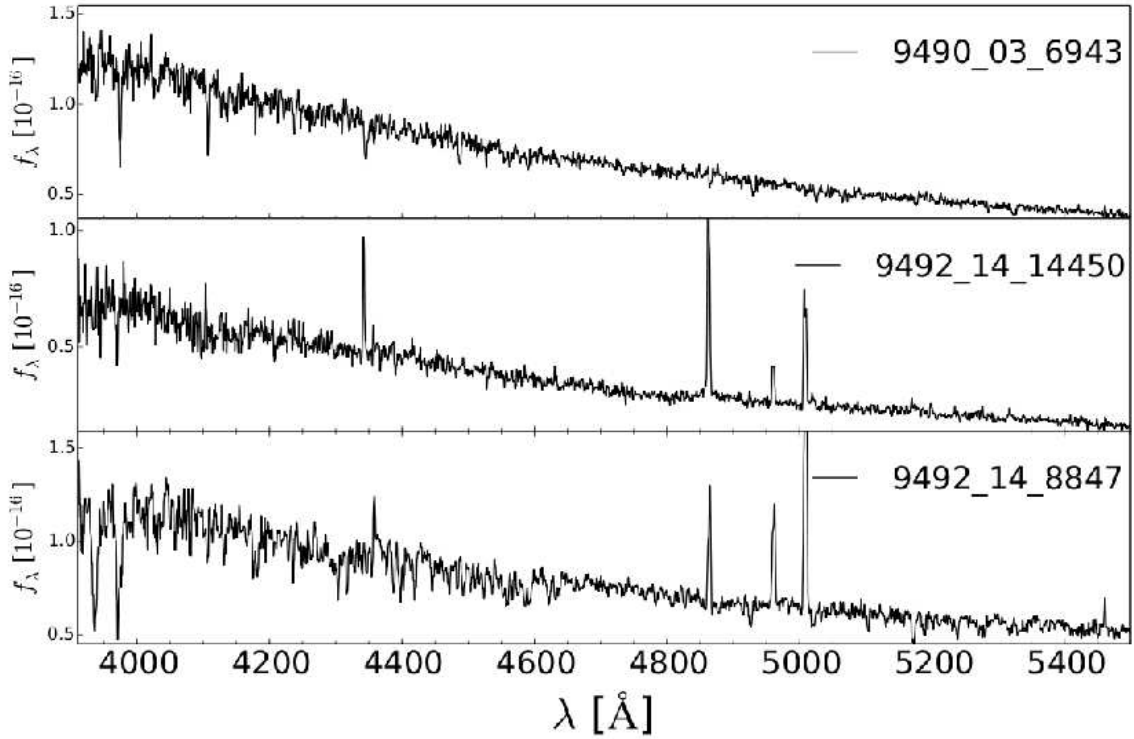}
\caption{Spectra of selected intermediate-type supergiants.The spectra are flux 
calibrated and smoothed with a 3 box smooth in the {\it splot} task in IRAF.}
\label{figIMSp}
\end{figure}

The spectrum  of \textit{9492$\_$14$\_$8847 (F5 I)}, in  Figure~\ref{figIMSp},  shows 
 strong  Ca II H and K lines and the luminosity sensitive  Fe II/Ti II blends at $\lambda\lambda$4172-4179.  The presence of a weak G-band indicates a spectral type  later than  $\approx$F2.  We estimate 9492$\_$14$\_$8847 to be approximately F5 based on  the absorption lines of Ca I $\lambda$4226, Fe I $\lambda$4046  and $\lambda$4383 lines, and Mn I $\lambda\lambda$4032. We were not able to use the Balmer lines in the classification due to strong nebular contamination. 
 
The $V$-band light curve for 9492$\_$14$\_$8847 (Figure~\ref{figSpecLC}) shows only 
minor variability  on the order of $\approx$0.1 magnitudes in amplitude.  The catalog photometry for 9492$\_$14$\_$8847 is $V$ = 19.69 and $(B-V)$ = 0.38.  The observed color is only marginally redder than expected, A$_{v}$ = 0.2 and  M$_{v}$ is -9.5 mag.  Adopting the bolometric correction for an F5 supergiant \citep[BC = 0.18;][]{Flower:1977, Flower:1996}, its bolometric luminosity of 9492$\_$14$\_$8847 is $M_{bol}$ = -9.3. Furthermore, the \textit{HST}/ACS $V$ image shows that 9492$\_$14$\_$8847 is on the periphery of a star-forming region (Figure A2) but appears to be a single point source.  Thus, we conclude that 9492$\_$14$\_$8847 is an intermediate supergiant which occupies the same part of the HR diagram as the yellow supergiants in M31 and M33 \citep{Humphreys:2014}.

The low S/N spectrum of \textit{9492$\_$14$\_$14450 (Early A I)} in Figure~\ref{figIMSp} shows emission lines of Fe II and [Fe II] and strong  H$\alpha$ and H$\beta$ emission with broad wings 
asymmetric to the red, indicative of Thompson scattering and the presence of a stellar 
wind. The relative strength of He I $\lambda$4471 and Mg II $\lambda$4481, suggest
an  early A spectral type. The [O III] lines and He I $\lambda$5876 are double, and $\lambda$5876 also has a P Cyg profile. While H$\beta$ and H$\gamma$ are not clearly 
double-peaked, they are both asymmetric to the red.

Based on the catalog photometry, 9492$\_$14$\_$14450 had an apparent magnitude of $V$ = 19.74  and an observed color of $(B-V)$ = 0.34 in January 2003.  The multi-epoch photometry from the LBT shows variability on the order of $\approx$0.4 magnitudes (Figure~\ref{figSpecLC}) with a maximum of $V$ = 19.95 in March 2009 (MJD 54912).   9492$\_$14$\_$14450 is located in a very crowded, star forming region (Figure A3), and based on its observed color has relatively high interstellar extinction, A$_{v}$ = 0.9 mag, assuming an A2 spectral type. M$_{v}$ is $\approx$ -10.2 mag and M$_{bol}$ is -10.3. Thus 9492$\_$14$\_$14450 has several characteristics of a stellar wind and given its high luminosity based on its catalog
photometry, it qualifies as a warm hypergiant star possibly similar to those in M31 and M33 \citep{Humphreys:2013}.

\textit{9490$\_$03$\_$6943} has an early to mid-A-type spectrum.. 
It is one of the few targets with no nebular contamination (Figure~\ref{figIMSp}). Its strong H$\alpha$ 
emission line with very broad wings and P Cygni absorption is thus circumstellar. 
It is relatively isolated (Figure A15) and appears to be a single star.The catalog
photometry is not contaminated and gives rather low visual extinction for a mid 
A spectral type ($\sim$ A2-A5) and M$_{v}$ = -9.0. 

The catalog photometry, visual extinction, spectral types, and any other notes regarding the remaining intermediate-type  supergiants are given in Table~\ref{tab:spec_members}.

\begin{figure}
\figurenum{5}
\epsscale{1.0}
\plotone{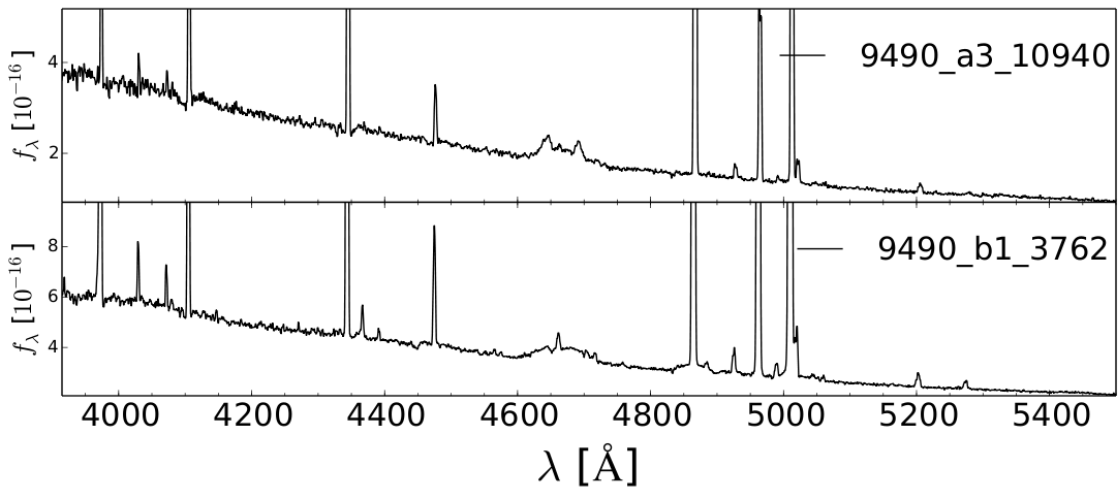}
\caption{Spectra of the two emission line stars.The spectra are flux calibrated 
and smoothed with a 3 box smooth in the {\it splot} task in IRAF.}
\label{figEM}
\end{figure}

\subsection{Emission-Line Sources}
The emission-line stars  have  a blue continuum and strong emission lines.  The distinction between targets in this group and the hot supergiants is the lack of absorption features permitting the estimation of a spectral type. There are only three stars in 
this group and one appears to be an H II region. 

The spectrum of \textit{9490$\_$a3$\_$10940(Of/WN)} is dominated by strong H and He I emission as well as the broad N III and N II emission features seen in WN stars (Figure~\ref{figEM}).  Some of the weaker emission lines include He II $\lambda$4686, O II $\lambda\lambda$4070-4076, and Fe II.  Based on its spectral features,  9490$\_$a3$\_$10940 is an Of/WN star.  The catalog magnitude for 9490$\_$a3$\_$10940 is $V = 21.49$ which is much fainter than our cutoff of $V \lesssim 20.5$ based on HLA photometry.  Inspection of the $V$-band ACS image (Figure A7) shows that 9490$\_$a3$\_$10940 is in a crowded 
region and the HLA aperture photometry is contaminated by nearby stars.  The observed color  is $(B-V) = -0.06$ which is approximately 0.2 magnitudes redder than we would expect for a star as hot as 9490$\_$a3$\_$10940.  Though we do not have a precise spectral type for 9490$\_$a3$\_$10940, adopting  $(B-V)$ $\approx$ -0.3 mag, which is typical for hot O-type stars,  the extinction towards 9490$\_$a3$\_$10940 is roughly 0.7-0.8 magnitudes and M$_{v}$ is -8.3 mag.  Correcting for extinction and adopting a bolometric correction of -3.0, we estimate the bolometric luminosity to be M$_{bol} \approx -11.3$.

 \textit{9490$\_$b1$\_$3762} is in the giant star-forming complex NGC 5461 (Figure A30) and many of the emission features in its spectrum  (Figure ~\ref{figEM}) 
 are nebular. Despite the strong nebular contamination, the H$\alpha$ and H$\beta$ line profiles have very broad wings suggesting a strong stellar wind.  We also 
 identify He I, C III $\lambda\lambda$4647-4652, and C IV $\lambda$4658 weak absorption lines indicating an underlying hot star.Emission lines of [S II]   Fe II, [Fe II], and [Fe III] and broad WN nitrogen emission region from  from 4620 - 4720{\AA} are present. As in several other stars, H$\beta$, the [O III] lines and He I $\lambda$5876 have  split emission profiles.

The catalog photometry for 9490$\_$b1$\_$3762 gives $V$ = 22.13 and $(B-V)$ = 0.23.  The emission lines from the highly-ionized species we see in the spectrum can only be produced by the radiation from an OB-type star with $(B-V) <$ 0.  
Since the observed color is much redder than we expect for an OB-type star, it is likely that 9490$\_$b1$\_$3762 is highly reddened or a blend of objects. Since it is in such a highly crowded region (Figure A30) we cannot determine its variability from the LBT images. 

The low S/N spectrum  of \textit{B4} \citep{Sandage:1983} has the emission line spectrum of an H II region. H$\beta$, He I $\lambda$5876, and the [O III] lines show split or
double profiles.

\subsection{Candidate LBVs}
A census of variable stars in M101 was first conducted by \cite{Sandage:1974c} who identified nine irregular blue variables (V1 - V9).  Subsequent photometric studies  confirmed the variable nature of V1 and V2, and added a tenth star \citep[V10;][]{Sandage:1983}.  Fifty-year historical light curves for the candidate LBVs V1, V2, and V10 are presented in \cite{Sandage:1983}.  V3 through V9 are known to be variable, and are considered to be candiate LBVs, but lack historical light curves.  Here we present the light curves and spectra for V2, V4, and V9.  We have also identified an additional  
candidate LBV, 9492$\_$14$\_$11998.  

\begin{figure}
\figurenum{6}
\epsscale{1.0}
\plotone{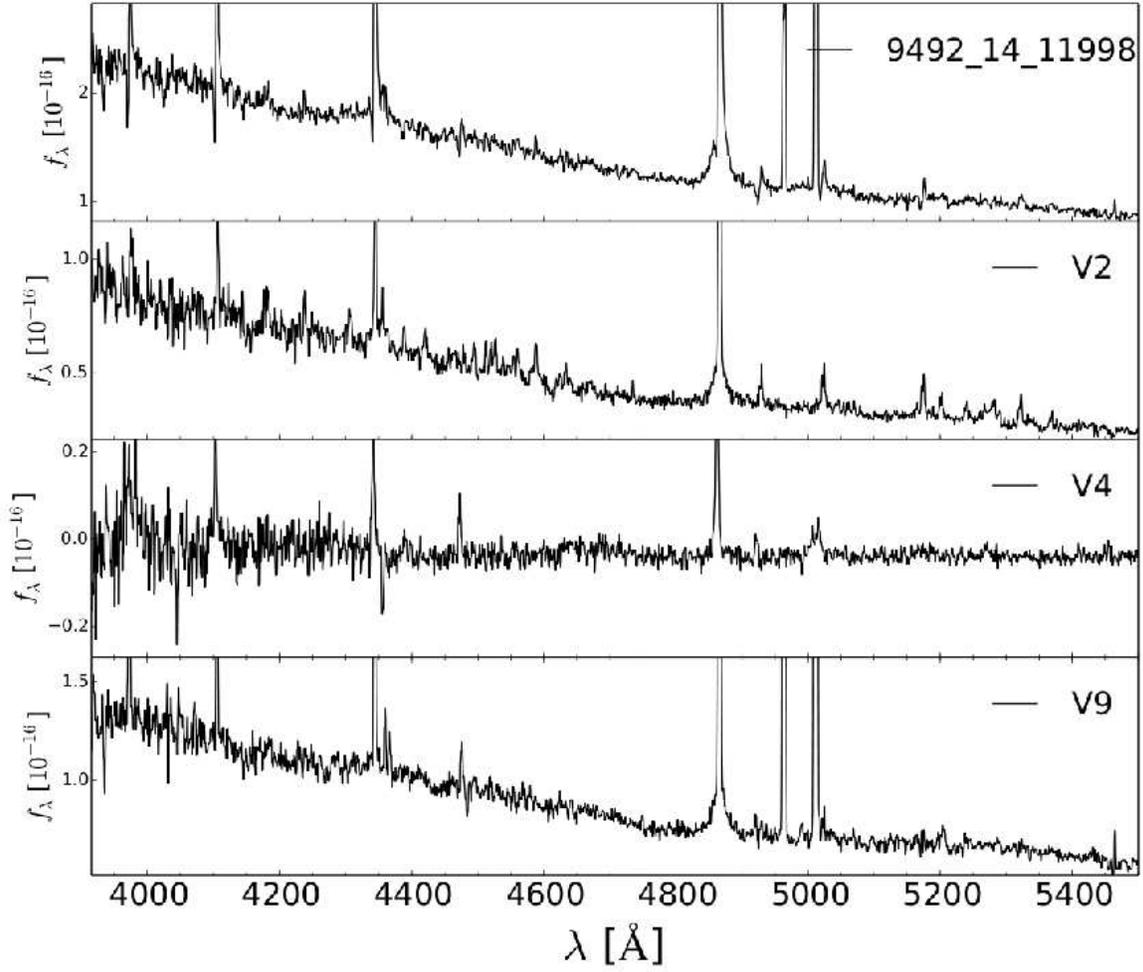}
\caption{Spectra of the four LBV candidates. The spectra are flux calibrated and
 smoothed with a 3 box smooth in the {\it splot} task in IRAF.}
 \label{figLBV}
 \end{figure}

The spectrum of \textit{9492$\_$14$\_$11998} shown in Figure~\ref{figLBV} is dominated by Balmer and He I emission lines with strong P-Cygni profiles. The line profiles of H$\alpha$ and H$\beta$ have very broad wings and are asymmetric to the red, a feature characteristic of Thompson scattering.  We estimate a wind velocity of $369\pm9$ km s$^{-1}$ from  the P Cygni absorption component in the hydrogen and He I emission lines. This is somewhat higher than the wind velocities of the M31 and M33 LBVs measured the same way \citep{Humphreys:2014}, but on the low end for normal OB supergiants \citep{Crowther:2006, Mokiem:2007}.  Weak Fe II emission is also present. We also identify absorption lines of He I at $\lambda\lambda$4009,4026,4121 and 4144{\AA}, and Si II at $\lambda$4128-31{\AA} typical of an early B2-B3-type supergiant 

The $V$-band light curve for 9492$\_$14$\_$11998 in Figure~\ref{figSpecLC}  shows that the star has steadily increased in brightness by approximately 0.2 magnitudes over the last 4 years.  The spectrum for 9492$\_$14$\_$11998 was obtained at its current visual maximum  of 18.9 mag.  The catalog photometry, which was observed in January 2003, indicates that  9492$\_$14$\_$11998 had a significantly fainter  apparent magnitude ($V$ = 19.40) than its present value.  Its observed color in 2003 was $(B-V)$ = 0.19 and based on our LBT photometry, it has not changed.

If we assume an intrinsic color of $(B-V)_{0} \approx -0.1$ corresponding to a mid-B-type supergiant \citep{Flower:1977, Flower:1996} and a Galactic extinction curve, the  visual extinction is $A_{V} \approx 0.9$ magnitudes.  Correcting the \textit{HST}/ACS visual magnitude for this extinction,  9492$\_$14$\_$11998 has an absolute visual magnitude of $M_{V} = -10.55$ magnitudes.  Adopting the bolometric correction (BC) for a mid-B supergiant \citep[BC $\approx$ -1.05;][]{Flower:1977, Flower:1996}, 9492$\_$14$\_$11998 has a high bolometric luminosity of $M_{bol}$ = -11.6, but  inspection of the \textit{HST}/ACS $V$ image (Figure A1) shows that it is in a crowded region and the photometry is likely contaminated by neighboring stars.  Despite the photometric uncertainties introduced by crowding, the spectral features of 9492$\_$14$\_$11998 are similar to an LBV in quiescence.  Continued monitoring will be necessary to determine whether or not 9492$\_$14$\_$11998 is  an LBV.

\textit{V2} is one of the three previously identified LBV candidates  in M101 \citep{Sandage:1983}.  \cite{Humphreys:1987} describe its spectrum as having H$\alpha$, H$\beta$, [O II] $\lambda$3727, and Fe II in emission.  Our higher S/N spectrum confirms the described features and reveals broad-winged H$\alpha$ and H$\beta$ emission lines, as well as  [N II], and [S II] in emission in the red. H$\beta$ and  the Fe II 
emission lines  redward of H$\beta$ have split-emission-line profiles.  These characteristics along with the fact that there is no [O III] $\lambda$4959 and $\lambda$5007 in emission suggests that the [N II], and [S II] lines may originate in the circumstellar environment rather than a nearby H\,{\small II} region.  Furthermore, the \textit{HST}/ACS $V$-band image (Figure A29) shows that V2 is a single star and there is no indication of an H\,{\small II} region within the fiber.

The 50-year historical light curve for V2 is shown in \cite{Sandage:1983}.  Over the 50 years it was monitored, V2  faded from $V \approx 19.1$ in 1910 to $V \approx 20.3$ in 1960.  Our 4.5-year $V$-band light curve (Figure~\ref{figSpecLC}) indicates that V2 is increasing in brightness from $V$ = 20.6 in 2008 to $V$ = 20.25 in 2013.  The current spectrum shows V2 to be a hot star and, therefore, in quiescence if it is an LBV.      

The spectrum of \textit{V4} is shown in Figure~\ref{figLBV} and is fairly noisy despite smoothing.  It is dominated by  strong H, He I, [N II], and [S II] emission lines.   The spectrum of V4 also shows Fe II emission throughout, although the lines are not particularly strong. 

Although V4 was originally identified as a variable star in M101, \citep{Sandage:1974c},
 there is no historical light curve.  The LBT light curve for V4 displays variability with 
 an  amplitude of approximately 0.4 magnitudes in the $V$-band (Figure~\ref{figSpecLC}) with a maximum magnitude of $V$ = 21.9. The catalog photometry from observations obtained in 2003 show V4 to be at $V$ = 22.0.  \cite{Sandage:1974c} list the visual minimum and maximum for V4 to be $V$ = 22.2 and $V$ = 19.9, respectively.  Assuming their photometry to be accurate, V4 is currently at its visual minimum and the oscillations we see are likely the low-amplitude variations commonly superposed on the longer time scale LBV minima and maxima.

The spectrum of \textit{V9} in Figure~\ref{figLBV} shows strong Balmer emission.  The H$\beta$ and H$\alpha$ lines have broad, asymmetric wings indicating a strong stellar wind.  We do not see any indication of P-Cygni profiles, however, this could be due to the large amount of nebular contamination filling in the absorption lines.   
Absorption lines of Fe II, Mg II $\lambda$4481, and the Ca II K-line are present. 
The luminosity sensitive O I $\lambda$ 7774 in intermediate type stars is also present.
 V9 is located on the periphery of a large star-forming complex which is not within the \textit{HST}/ACS footprint.  However, we have examined the LBT images (Figure A4) and although the complex is not resolved, V9 is located far enough away that its spectrum and photometry should  not be seriously contaminated, and  it is unlikely that the spectrum is a blend 

The LBT $V$-band light curve for V9 indicates that the star has increased in visual brightness by approximately 0.8 magnitudes over the last 4 years from  $V$ = 20.2 in 2008 to $V$ = 19.6 in 2012. The most recent data suggest V9 may be fading again.  
Similar to V4, V9 was originally identified as a variable star in M101  
\citep{Sandage:1974c} with similar  
minima and maxima: $V_{min}$ = 20.3 (1950) and $V_{max}$ = 19.5 (1927/1947). When the Hectospec spectrum was observed, V9 was near its visual maximim based on the LBT light 
curve (Figure~\ref{figSpecLC}). Although dominated by a strong continuum and Balmer emission, its spectrum suggests that the underlying star was an early A-type supergiant and therefore
may be an LBV in its optically thick wind  or visual maximum  state. Given its spectrum and variability history, it is very likely that V9 is an LBV.      

\section{The Variables Without Spectra}
In this section, we discuss the light curves for the targets without spectra that met our criterion for variability.  We separate the variables by their photometry into three groups: hot or early-type (O$-$B), intermediate (A$-$F), and cool (G$-$M) stars.  Assuming zero reddening, the selection criteria for our groups corresponds to $(B-V) < 0$, $0 < (B-V) < 0.9$, and $(B-V) > 0.9$, respectively.  Based on the our estimates of A$_{V}$ for our spectroscopically confirmed members (Table~\ref{tab:spec_members}), the majority of the stars are likely to be at least somewhat reddened.  Therefore, we have relaxed our selection criteria to: hot [$(B-V) \leq 0.2$], intermediate [$0.2 < (B-V) < 1.0$], and cool [$(B-V) > 1.0$].   When available, we used the \textit{HST}/ACS catalog 
photometry to assign the stars to the three groups.  For targets that were not recovered in the catalog, we used the LBT photometry, although   the LBT colors are unlikely to be precise in crowded regions.  

We have LBT spectra for many of the stars in this section.  When the improved
 reduction pipeline is complete,  we  will discuss the spectra along with 
 their light curves in a future paper.

\begin{figure}
\figurenum{7}
\epsscale{0.6}
\plotone{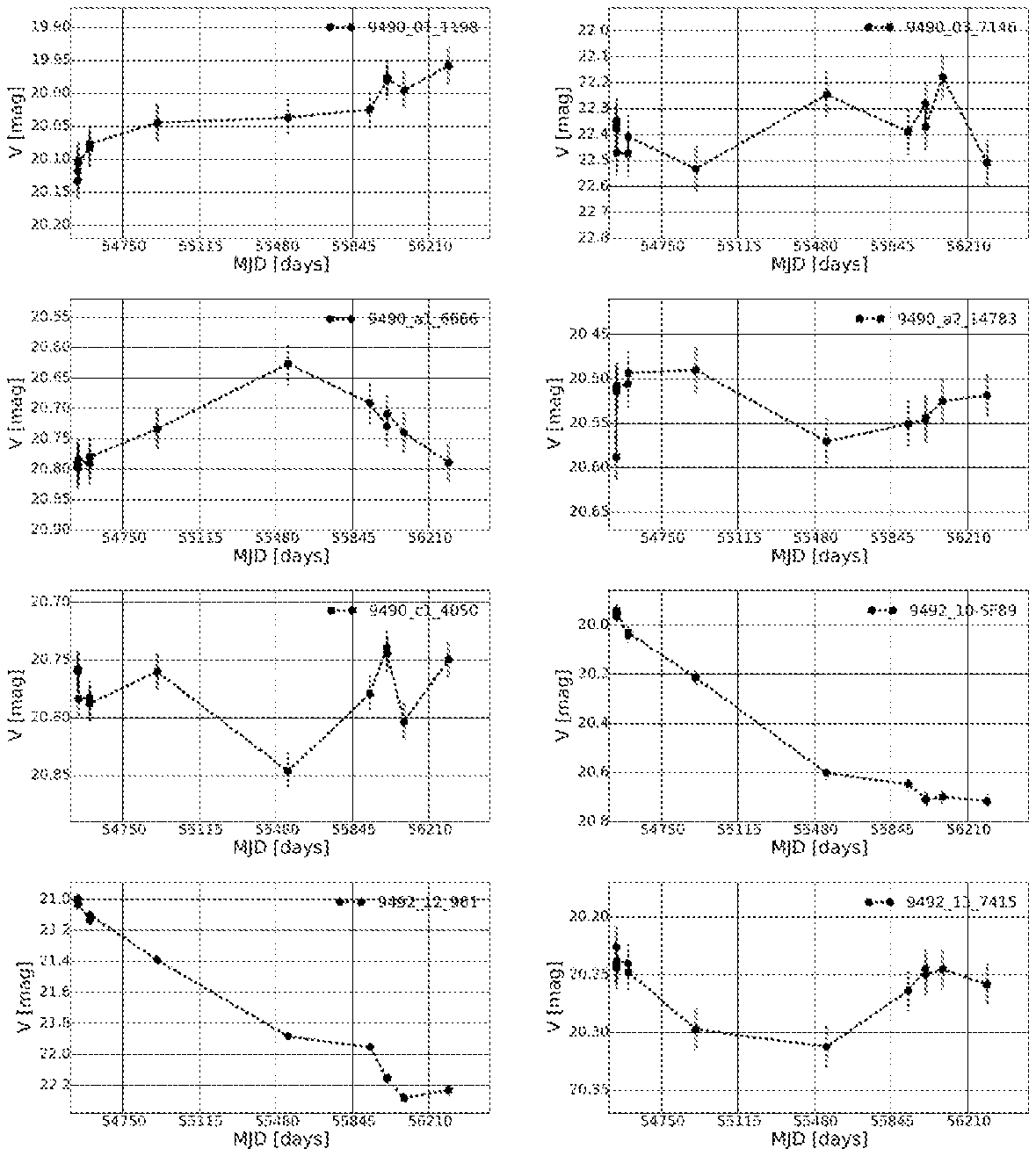}
\caption{The $V$-band light curves for the early-type  variables (i.e. targets t
hat meet our criterion for variability
and have $(B-V) \leq 0.2$).  The x-axis grid lines represent 365 days.}
\label{figBlueLC}
\end{figure}

\subsection{Early-Type  Variables}
Figure~\ref{figBlueLC} shows the $V$-band light curves for the eight early-type  variables.  The majority have photometric variability on the order of a few tenths of a magnitude.  Two of the stars, 9492$\_$10$\_$SF89 and 9492$\_$12$\_$961, show changes in $V$-band that are $\approx$1 magnitude.  In both cases, the stars are fading.  In 2003, they  had $V$-band magnitudes of $V$ = 21.22 and $V$ = 20.94, respectively, measured  from the \textit{HST}/ACS images.  In March 2008, the LBT $V$-band magnitudes did not differ by more than a tenth of a magnitude compared to 2003.  Though we do not have data between 2003 and 2008, the similarities suggest that their  decline in brightness may be a recent development or alternatively the stars may be semi-periodic.  Given their photometric colors and variability, 9492$\_$10$\_$SF89 and 9492$\_$12$\_$961 may also be LBVs.  Spectral analysis and photometric monitoring will be necessary for confirmation.

\subsection{Intermediate-Type  Variables}
\begin{figure}
\figurenum{8}
\epsscale{0.6}
\plotone{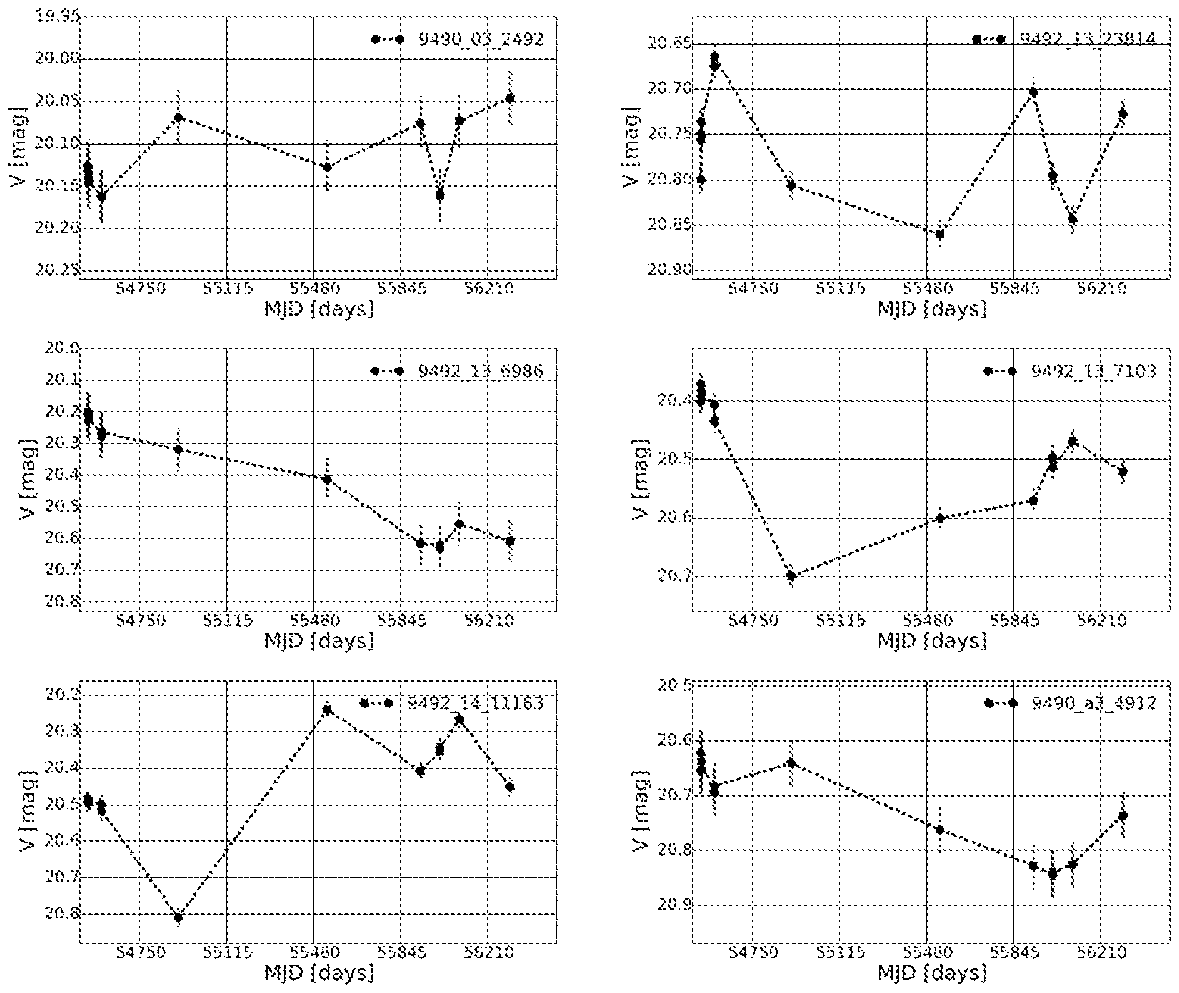}
\caption{The $V$-band light curves for the intermediate-type  variables (i.e. ta
rgets that meet our criterio
n for variability and have $0.2 < (B-V) \leq 1.0$).  The x-axis grid lines repre
sent 365 days.}
\label{figImLC}
\end{figure}

The $V$-band light curves for the six variables with $0.2 < (B-V) \leq 1.0$ are shown  in Figure~\ref{figImLC}.  Our color criteria for the intermediate group could include supergiant Cepheids.  To ensure that we have not re-identified any known Cepheids, we have cross-referenced the astrometry of the intermediate variables with the \cite{Shappee:2011} catalog of M101 Cepheids and removed any matches.  

The intermediate-type  variables have photometric properties that overlap with the intermediate type  supergiants and with LBVs at maximum light \citep{Humphreys:1994}.  It is typical for A to F supergiants to exhibit variability on the order $0.1-0.2$ magnitudes known as $\alpha$ Cygni variability \citep{van-Genderen:2002}.  LBVs can also exhibit $\alpha$ Cygni variability, during an extended maximum,  but with a  higher amplitude \citep[$\pm0.5$ magnitudes;][]{van-Genderen:1997a, van-Genderen:1997b}.  The variables presented here primarily exhibit $\alpha$ Cygni variability typical of intermediate-type  supergiants.  
However, 9492$\_$13$\_$6986 and 9492$\_$13$\_$11163 show larger amplitude variability 
on the order of $\approx$0.5 magnitudes.  Furthermore, 9492$\_$13$\_$11163 was brighter by 0.5 magnitudes in January 2003 from the \textit{HST}/ACS  catalog photometry compared with the  March 2008 LBT $V$-band magnitude.  Without a spectrum, we cannot say with any 
certainty that 
these stars are not typical intermediate-type supergiants, however, their variability suggests they could be LBVs. Followup observations will be necessary for confirmation. 

\begin{figure}
\figurenum{9}
\epsscale{0.6}
\plotone{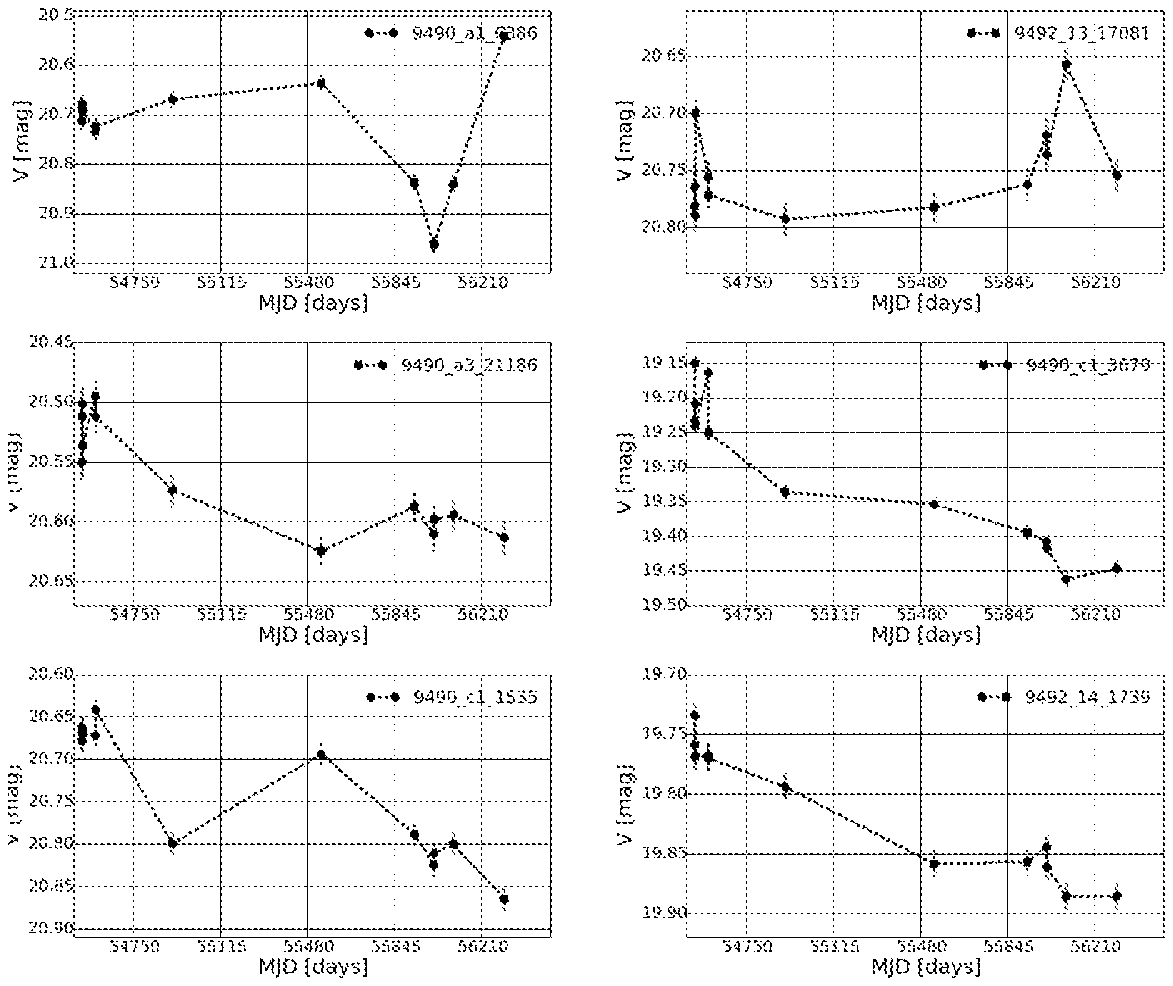}
\caption{The $V$-band light curves for the cool variables (i.e. targets that mee
t our criterion for variability and have $(B-V) > 1.0$).  The x-axis grid lines represent 365 days.}
\label{figRedLC}
\end{figure}

\subsection{Cool Variables}
Finally, we present the $V$-band light curves (Figure \ref{figRedLC}) for the six stars which meet our variability criterion and have $(B-V) > 1.0$.  This group very likely  includes  red supergiants and foreground K and M dwarfs.  The stars in this group exhibit variability in the $V$-band that is $0.2-0.4$ magnitudes in amplitude which is typical of K and M supergiants \citep[and references therein]{Meynet:2011}, but based on the analysis of \cite{Grammer:2013a}, it is likely that a large fraction of the cool variables are foreground.  We present the light curves here, but defer the analysis to a future paper which will include the spectra.

\section{Summary and Future Work}

\begin{figure}
\figurenum{10}
\epsscale{0.8}
\plotone{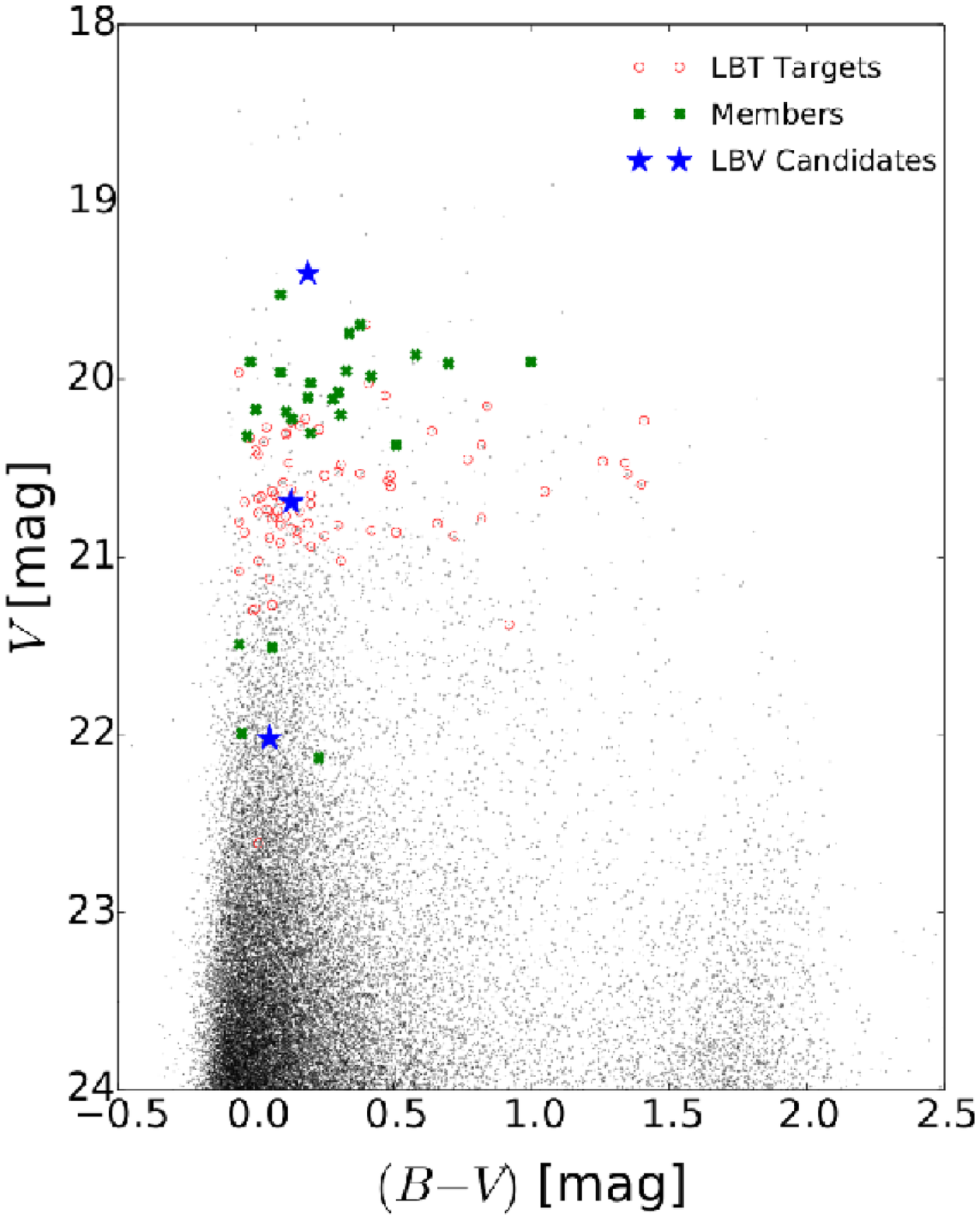}
\caption{The color-magnitude diagram  from Paper I showing the confirmed members
, the LBVs, and the LBT targets.  Confirmed non-memebers have
been removed.}
\label{figCMD}
\end{figure}

In this third paper on our survey of the massive star population in M101, we present
the results of spectral classification and multicolor photometry of 50 of the 
visually brightest stars in the field and confirm  that 31 are members of M101. 
It is not 
surprising that the majority are intermediate-type supergiants with A to F-type
spectra since we selected the targets based on their visual apparent
magnitude and variability. We present new photometry and light curves for three
candidate LBVs, V2, V4 and V9 and identify a new candidate, 9492$\_$14$\_$11998.
Their spectra and variability confirm that they are LBV candidates and V9 may be in
an LBV-like maximum light state or eruption.  Followup spectra and continued 
photometric monitoring  of these candidates will be necessary for confirmation 
and to determine the nature of their observed variability.  
We also discuss the light curves of 20 variables that lack spectroscopy. 
Four  of these stars with large amplitude variability may be LBV candidates.  

Figure 10 shows the observed color-magnitude diagram from Paper I 
with the confirmed members identified. The LBV candidates and the  targets
observe with the LBT are identified separately and the non-members have been removed.

Forty-six additional stars have been observed with the LBT/MODS. 
When the LBT/MODS reduction pipeline is completed, the addition of these stars including many of the variables will allow a more complete sampling of 
the upper HR Diagram in M101.

\acknowledgements
Research by R. Humphreys and S. Grammer  on massive stars is supported by  
the National Science Foundation grant AST-1109394. Based on observations made 
with the NASA/ESA Hubble Space Telescope, and obtained from the Hubble Legacy Archive, which is a collaboration between the Space Telescope Science Institute (STScI/NASA), the Space Telescope European Coordinating Facility (ST-ECF/ESA) and the Canadian Astronomy Data Centre (CADC/NRC/CSA).

{\it Facilities:} \facility{MMT/Hectospec, LBT/LBC, HST/ACS}

\appendix
\section{Snaphot images}

Snaphot images of 30 of the confirmed members from the {\it HST}/ACS F555W  frame  are shown here with a 1$\farcs$5 diameter circle equal to the Hectospec aperture. At the distance of M101, this is 47 pc. The images  are displayed in the same order as the stars in Table 1.  There is no image for B4. 
The snapshots are displayed here in pairs to save space. In the published on-line paper thay will be separate and interested readers will be able to click on 
each for an enlarged view. 

\begin{figure}
\figurenum{A1,2}
\epsscale{0.4}
\plottwo{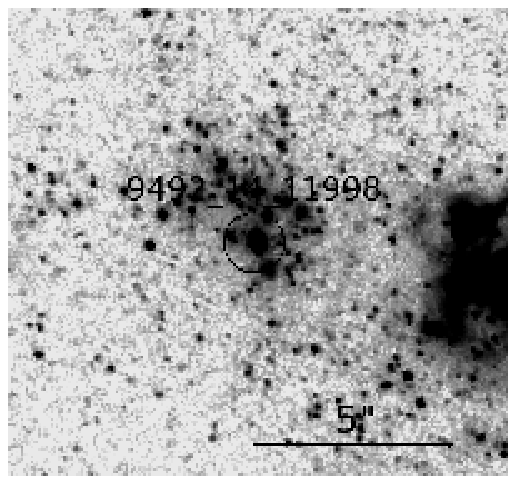}{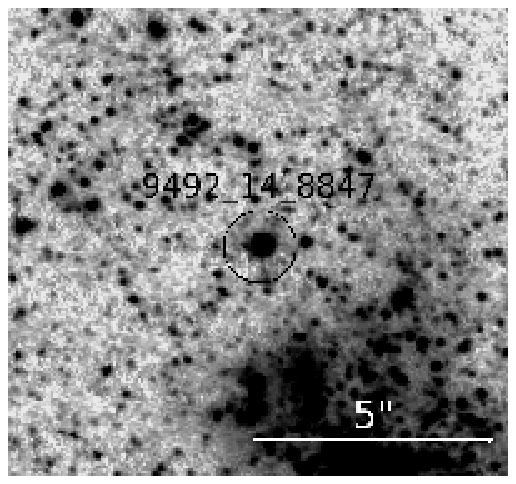}
\caption{a. 9492$\_$14$\_$11998 b. 9492$\_$14$\_$8847}
\end{figure}

\begin{figure}
\figurenum{A3,4}
\epsscale{0.4}
\plottwo{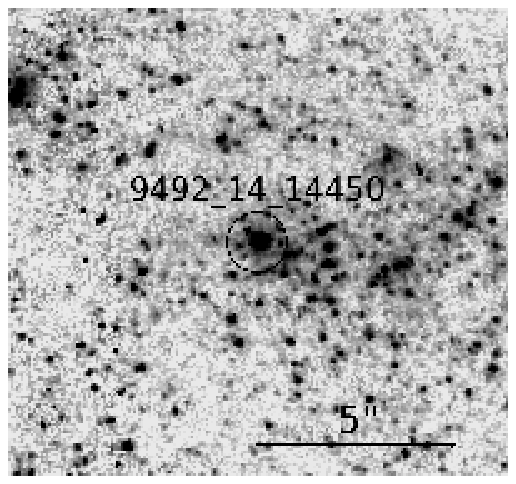}{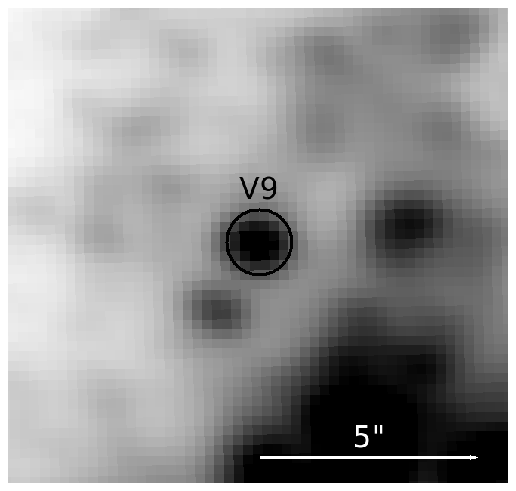}
\caption{a. 9492$\_$14$\_$14450 b. V9, an LBV candidate, LBT image.}
\end{figure}

\begin{figure}
\figurenum{A5,6}
\epsscale{0.4}
\plottwo{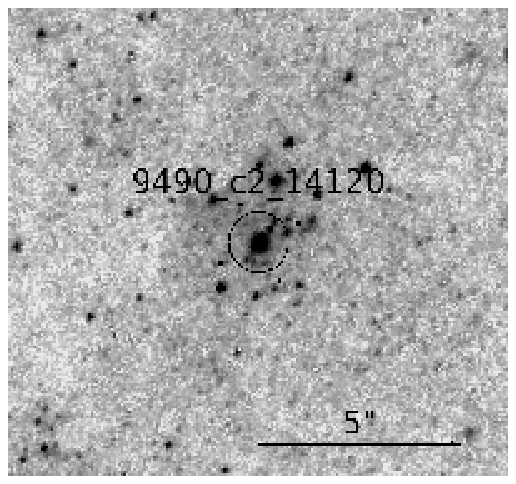}{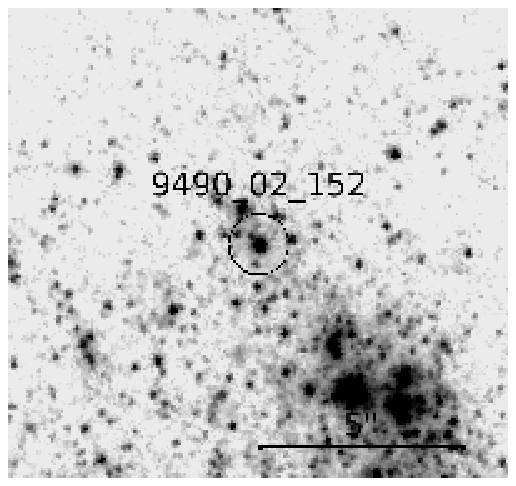}
\caption{a. 9490$\_$c2$\_$14120 b. 9490$\_$02$\_$152}
\end{figure}

\begin{figure}
\figurenum{A7,8}
\epsscale{0.4}
\plottwo{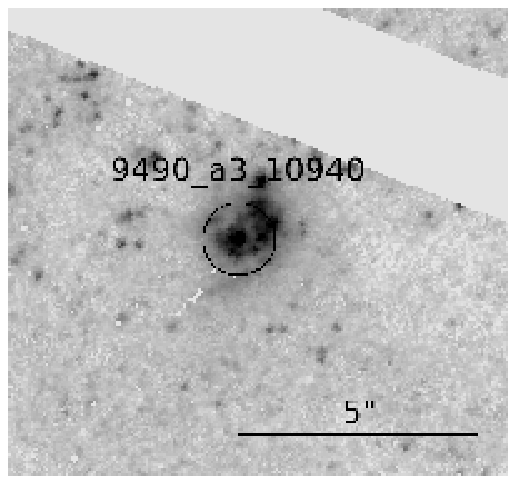}{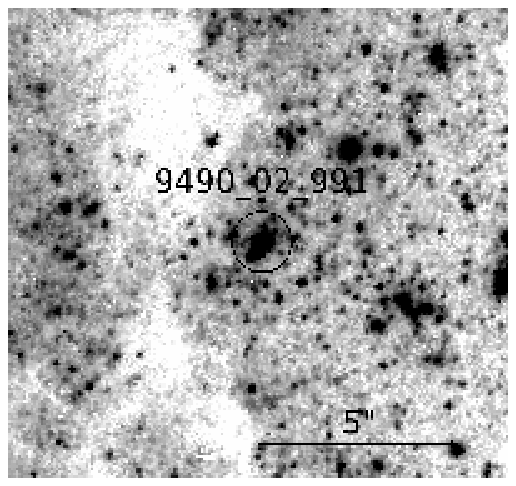}
\caption{a. 9490$\_$a3$\_$10940  b. 9490$\_$02$\_$991}
\end{figure}

\begin{figure}
\figurenum{A9,10}
\epsscale{0.4}
\plottwo{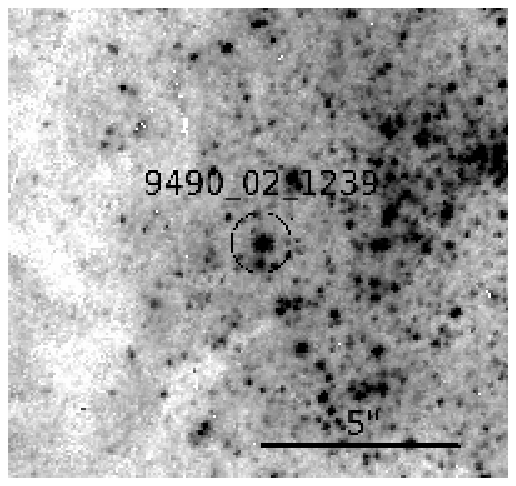}{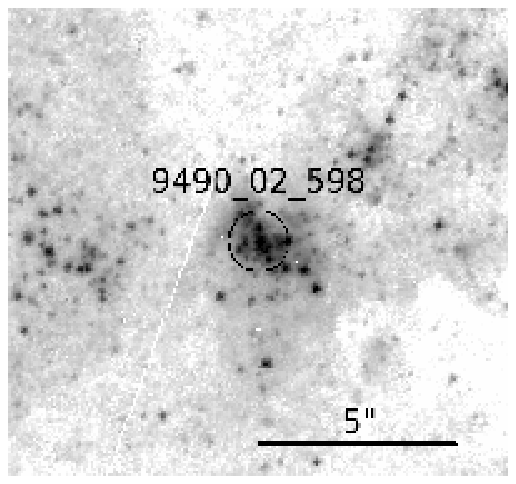}
\caption{a. 9490$\_$02$\_$1239 b. 9490$\_$02$\_$598}
\end{figure}

\begin{figure}
\figurenum{A11,12}
\epsscale{0.4}
\plottwo{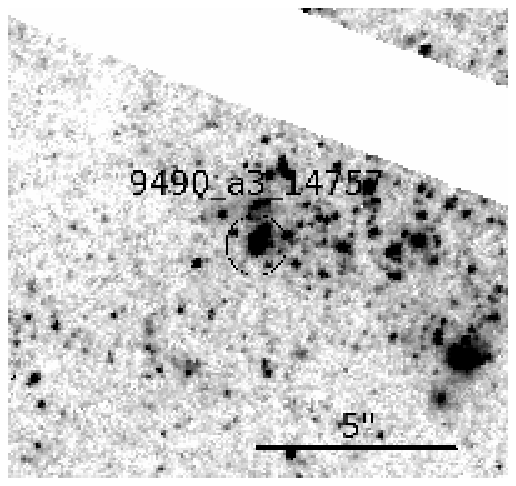}{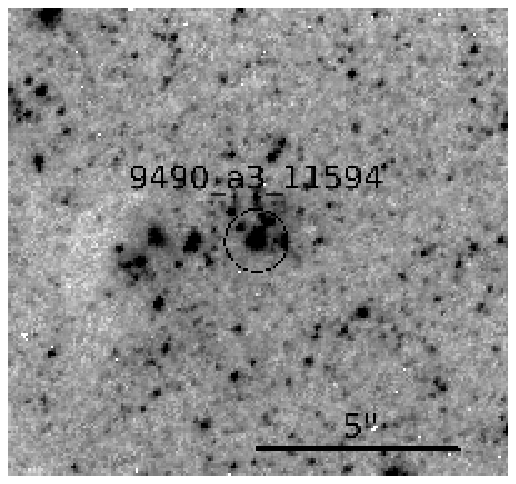}
\caption{a. 9490$\_$a3$\_$14757 b. 9490$\_$a3$\_$11594}
\end{figure}

\begin{figure}
\figurenum{A13,14}
\epsscale{0.4}
\plottwo{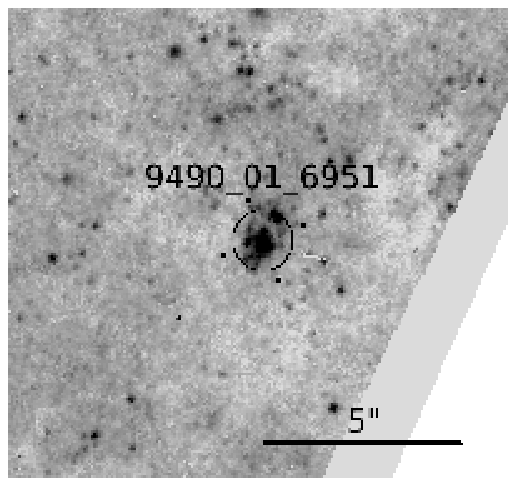}{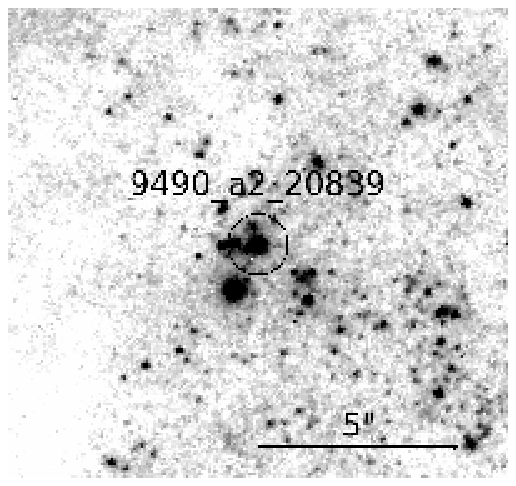}
\caption{a. 9490$\_$01$\_$6951  b. 9490$\_$a2$\_$20839}
\end{figure}

\begin{figure}
\figurenum{A15,16}
\epsscale{0.4}
\plottwo{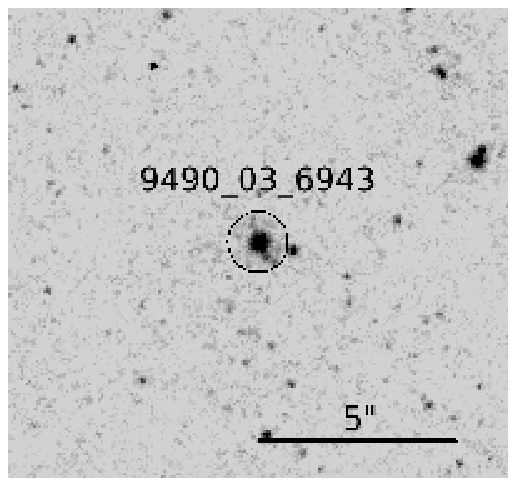}{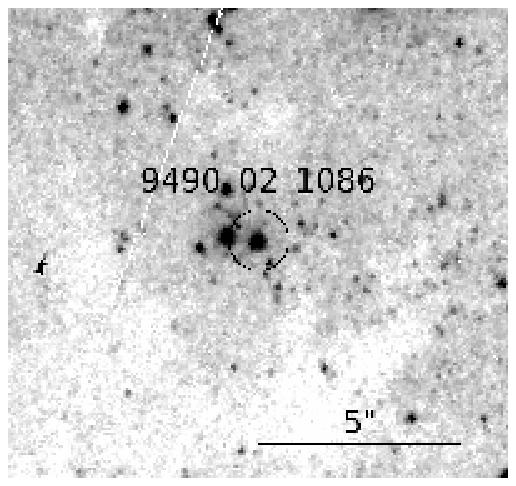}
\caption{a. 9490$\_$03$\_$6943  b. 9490$\_$02$\_$1086}
\end{figure}

\begin{figure}
\figurenum{A17,18}
\epsscale{0.4}
\plottwo{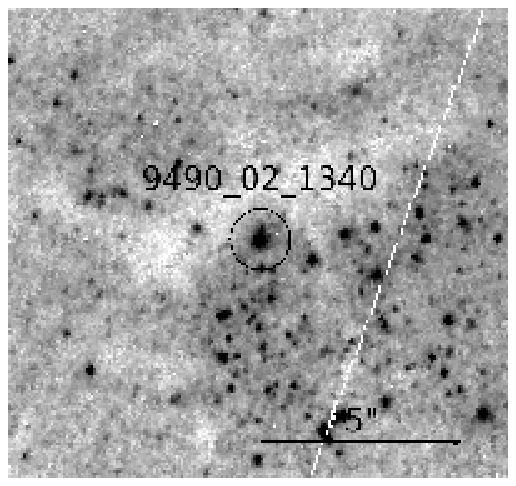}{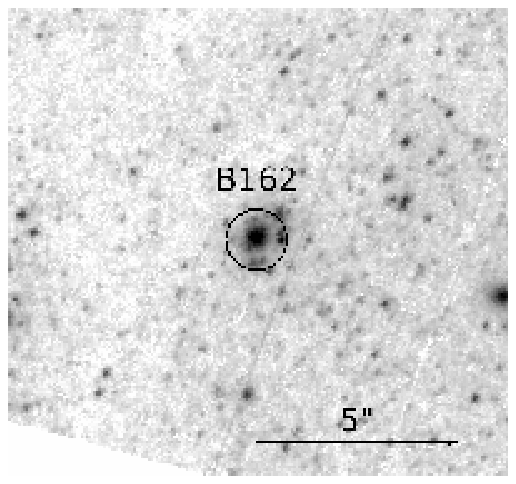}
\caption{a. 9490$\_$02$\_$1340 b. B162 }
\end{figure}

\begin{figure}
\figurenum{A19,20}
\epsscale{0.4}
\plottwo{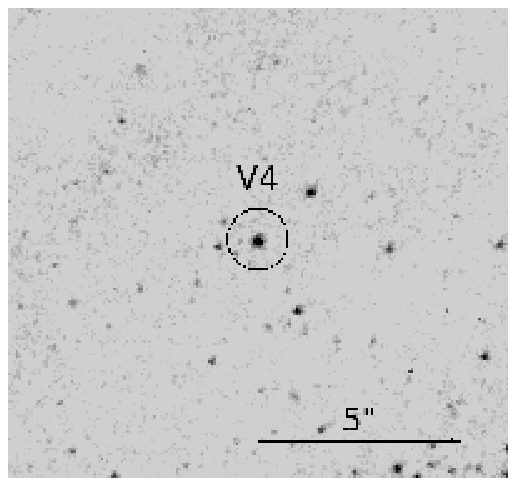}{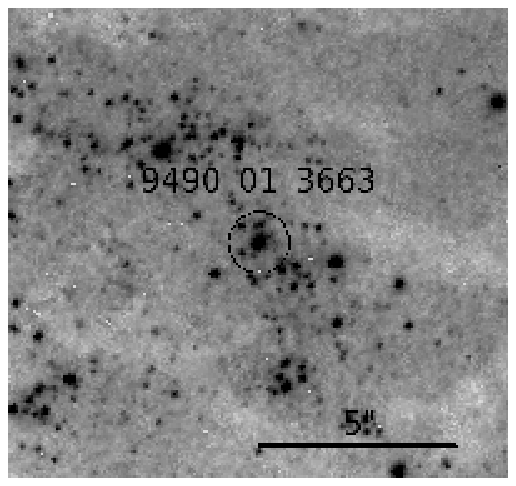}
\caption{a. V4, an LBV candidate. b. 9490$\_$01$\_$3663}
\end{figure}

\begin{figure}
\figurenum{A21,22}
\epsscale{0.4}
\plottwo{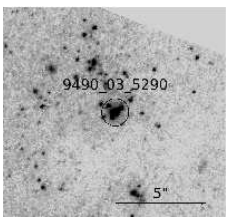}{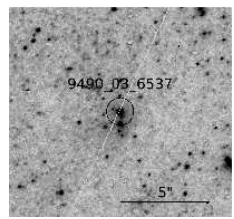}
\caption{a. 9490$\_$03$\_$5290 b. 9490$\_$03$\_$6537}
\end{figure}

\begin{figure}
\figurenum{A23,24}
\epsscale{0.4}
\plottwo{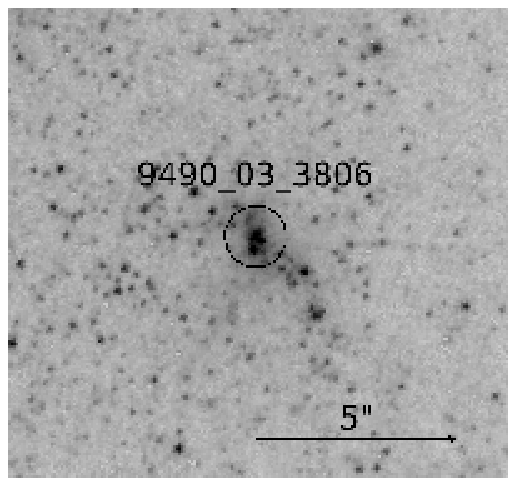}{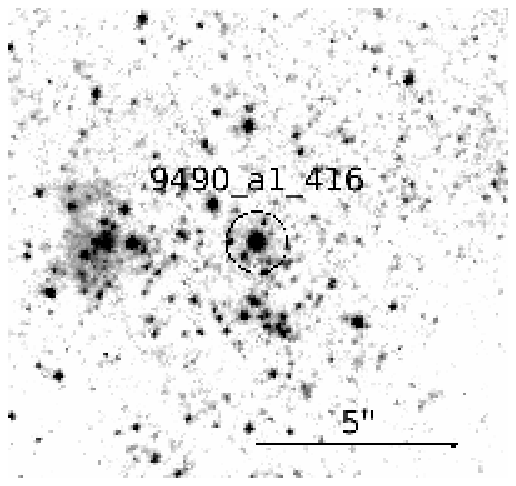}
\caption{a. 9490$\_$03$\_$3806 b. 9490$\_$a1$\_$416}
\end{figure}

\begin{figure}
\figurenum{A25,26}
\epsscale{0.4}
\plottwo{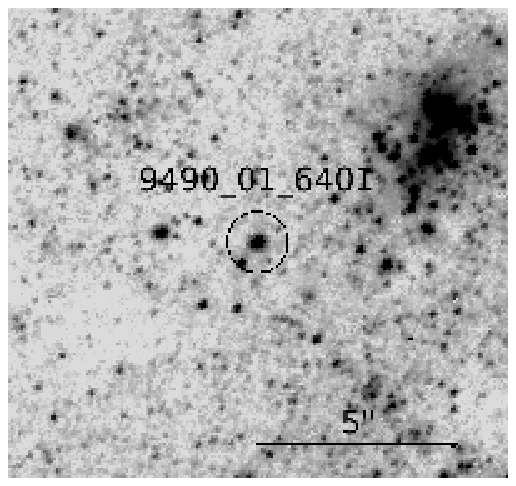}{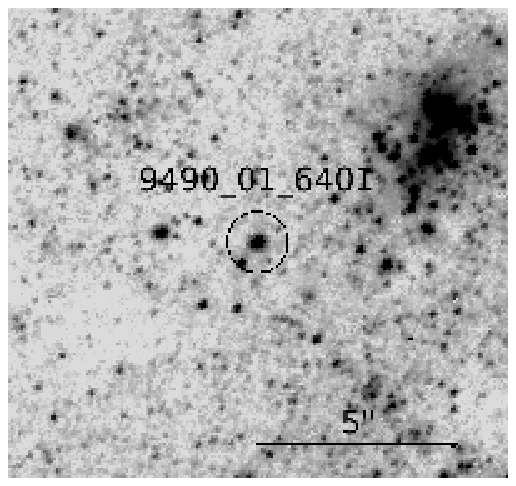}
\caption{a. 9490$\_$a1$\_$6401 b. 9490$\_$03$\_$1487}
\end{figure}

\begin{figure}
\figurenum{A27,28}
\epsscale{0.4}
\plottwo{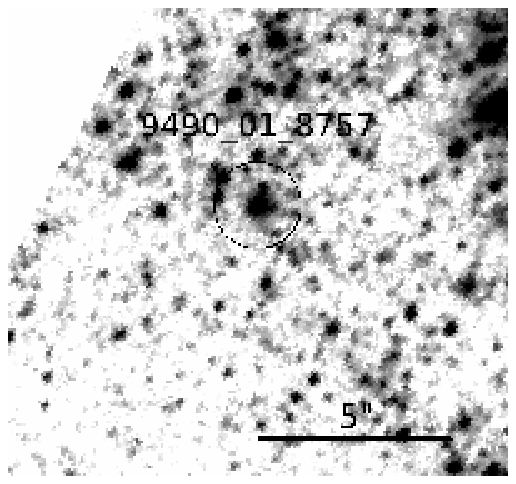}{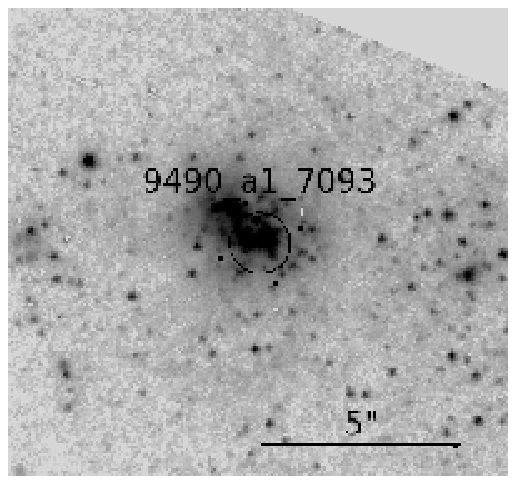}
\caption{a. 9490$\_$01$\_$8757 b. 9490$\_$a1$\_$7093}
\end{figure}

\begin{figure}
\figurenum{A29,30}
\epsscale{0.4}
\plottwo{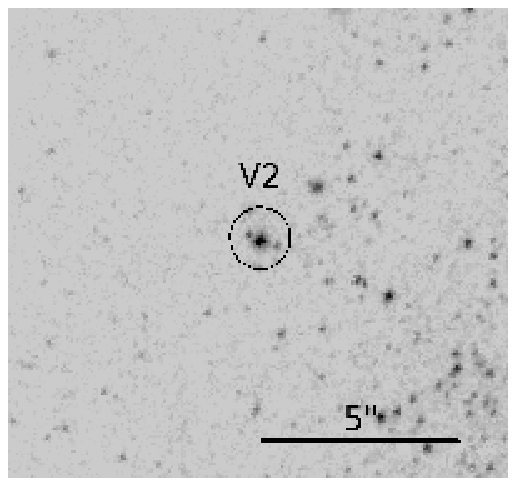}{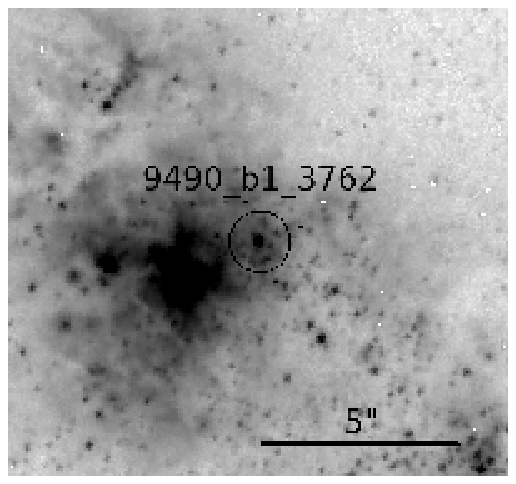}
\caption{a. V2, an LBV candidate. b. 9490$\_$b1$\_$3762}
\end{figure}

\clearpage 



\input{table1.tex}

\input{table2.tex}

\input{table3.tex}

\end{document}

%% file: table1.tex
\begin{deluxetable}{llcccccccl}
\tabletypesize{\scriptsize}
\rotate
\tablenum{1}
\tablecolumns{10}
\tablecaption{Members of M101}
\tablewidth{0pc}
\label{tab:spec_members}
\tablehead{\colhead{Catalog ID} & \colhead{Star ID} & \colhead{$\alpha_{J2000}$} & \colhead{$\delta_{J2000}$} & \colhead{$V$} & \colhead{$(B-V)$} & \colhead{Group} & \colhead{Variability} & \colhead{A$_{V}$} & \colhead{Comments}}

\startdata
  J140220.98+542004.38 &  9492$\_$14$\_$11998  &  14:02:20.98  &  54:20:04.38 & 19.40 & 0.19  & LBV Cand.  & $BVR$  & $0.9$  & P-Cyg H, He I; Fe II em.; see text, Figure 6\\
  J140227.30+541952.50 &  9492$\_$14$\_$8847   &  14:02:27.30  &  54:19:52.50 & 19.69 & 0.38  & Im. SG     & $BVR$  & 0.2        & F5 I, see text, Figure 4             \\
  ---                  &  B4                   &  14:02:27.89  &  54:16:18.43 & ---   & ---   & Em. Line   & ---    & ---        & Em. Line; H II region, see text\tablenotemark{c} \\
  J140228.83+542014.03 &  9492$\_$14$\_$14450  &  14:02:28.83  &  54:20:14.03 & 19.74 & 0.34  & Im. SG     & $UBVR$ & 0.9        & Early A I; see text\tablenotemark{c}, Figure 4 \\
  ---                  &  V9                   &  14:02:29.92  &  54:16:19.91 & ---   & ---   & LBV Cand.  & $UBVR$ & ---        & See text, Figure 6 \\
  J140248.46+541935.80 &  9490$\_$c2$\_$14120  &  14:02:48.46  &  54:19:35.80 & 20.10 & 0.19  & Im. SG     & ---    & $0.2-0.3$  & Late A I\\
  J140256.71+541834.09 &  9490$\_$02$\_$152    &  14:02:56.71  &  54:18:34.09 & 20.30 & 0.20  & Im. SG     & ---    & 0.2        & A5-F0  I \\
  J140259.37+542323.89 &  9490$\_$a3$\_$10940  &  14:02:59.37  &  54:23:23.89 & 21.49 & -0.06 & Em. Line   & ---    & $0.7-0.8$  & Of/WN; see text, Figure 5 \\
  J140301.84+541949.62 &  9490$\_$02$\_$991    &  14:03:01.84  &  54:19:49.62 & 20.17 & 0.00  & Hot SG     & ---    & 0.5        & early B-type, weak WN  \\
  J140302.61+542001.39 &  9490$\_$02$\_$1239   &  14:03:02.61  &  54:20:01.28 & 20.02 & 0.20  & Im. SG     & ---    & $\approx$0 & Early A I  \\
  J140304.70+541925.00 &  9490$\_$02$\_$598    &  14:03:04.70  &  54:19:25.00 & 20.07 & 0.30  & Hot SG     & ---    & 1.6        & early B I; see text\tablenotemark{c}, Figure 3 \\
  J140305.13+542342.14 &  9490$\_$a3$\_$14757  &  14:03:05.13  &  54:23:42.14 & 20.32 & -0.03 & Hot SG     & ---    & 0.3        & early B + WN, double emission profiles \\
  J140307.95+542326.81 &  9490$\_$a3$\_$11594  &  14:03:07.95  &  54:23:26.81 & 21.51 & 0.06  & Hot SG     & ---    & 0.8        & early B I; see text\tablenotemark{c}, Figure 3  \\
  J140309.19+542138.77 &  9490$\_$01$\_$6951   &  14:03:09.19  &  54:21:38.77 & 19.96 & 0.09  & Hot SG     & ---    & 0.3        & late B:; weak WN feature; composite. spectrum \\
  J140311.06+541830.96 &  9490$\_$a2$\_$20839  &  14:03:11.06  &  54:18:30.96 & 19.95 & 0.33  & Im. SG     & $V$    & 0.7        & A5  I \\
  J140311.32+542518.55 &  9490$\_$03$\_$6943   &  14:03:11.32  &  54:25:18.55 & 20.18 & 0.11  & Im. SG     & ---    & 0.2        & early - mid A\tablenotemark{a}, see text, Figure 4\\
  J140313.44+541954.44 &  9490$\_$02$\_$1086   &  14:03:13.44  &  54:19:54.44 & 19.90 & -0.02 & Hot SG     & ---    & 0.1        & late B,  H em, H$\beta$ double. \\
  J140313.74+542004.56 &  9490$\_$02$\_$1340   &  14:03:13.74  &  54:20:04.56 & 20.11 & 0.28  & Im. SG     & ---    & 0.3        & F0-F2 I; H em. \\
  J140314.80+541737.93 &  B162                 &  14:03:14.80  &  54:17:37.93 & 19.52 & 0.09  & Hot SG     & ---    & 0.4        & B8 I; see text, Figure 3 \\
  J140314.98+541645.26 &  V4                   &  14:03:14.98  &  54:16:45.26 & 22.02 & 0.05  & LBV Cand.  & $UBVR$   & ---        & See text, Figure 6 \\
  J140316.64+542042.04 &  9490$\_$01$\_$3663   &  14:03:16.64  &  54:20:42.04 & 21.99 & -0.05 & Hot SG     & ---    & 0.1        & B5 I, H$\beta$ double \\
  J140322.26+542437.69 &  9490$\_$03$\_$5290   &  14:03:22.27  &  54:24:37.62 & 19.98 & 0.42  & Im. SG     & ---    & 0.9        & A8 I \\
  J140323.43+542504.84 &  9490$\_$03$\_$6537   &  14:03:23.44  &  54:25:04.76 & 20.37 & 0.51  & Im. SG     & ---    & 1.2        & Late A I\\
  J140326.37+542411.63 &  9490$\_$03$\_$3806   &  14:03:26.37  &  54:24:11.63 & 20.22 & 0.13  & Hot SG     & ---    & $0.6-0.9$  & Early B I\tablenotemark{c}; weak WN feature \\
  J140328.35+541707.58 &  9490$\_$a1$\_$416    &  14:03:28.35  &  54:17:07.58 & 19.90 & 1.00  & Im. SG     & ---    & $\approx1.5$  & F: I \\
  J140328.86+542128.94 &  9490$\_$a1$\_$6401   &  14:03:28.86  &  54:21:28.94 & 19.91 & 0.70  & Im. SG     & ---    & $1.5-1.8$  & Late A I \\
  J140330.73+542335.77 &  9490$\_$03$\_$1487   &  14:03:30.73  &  54:23:35.66 & 19.86 & 0.58  & Im. SG     & ---    & $1.4$        & A5 I \\
  ---                  &  9490$\_$01$\_$8757   &  14:03:32.28  &  54:22:09.19 & ---   & ---   & Im. SG     & $V$    & ---        & Late F I\tablenotemark{b} \\
  J140332.78+542009.89 &  9490$\_$a1$\_$7093   &  14:03:32.78  &  54:20:09.89 & 20.20 & 0.31  & Hot SG     & ---    & $1.3$  & Mid B: I; see text\tablenotemark{c}, Figure 3\\
  J140332.88+542425.99 &  V2                   &  14:03:32.88  &  54:24:25.99 & 20.69 & 0.13  & LBV Cand.  & $UBVR$ & ---        & See text\tablenotemark{c}, Figure 6 \\
  J140341.18+541905.30 &  9490$\_$b1$\_$3762   &  14:03:41.18  &  54:19:05.30 & 22.13 & 0.23  & Em. Line   & ---    & ---        & Strong WN feature, see text\tablenotemark{c} \\
\enddata
\tablenotetext{a}{The residual [O I] night sky lines in the red appear to have P Cygni profiles due to poor sky subtraction.} 
\tablenotetext{b}{The nebular [N II] and [S II] lines appear to have have P Cygni profiles due to poor background/sky substraction.}
\tablenotetext{c}{Double or split emission lines.}
\end{deluxetable}

%% file: table2.tex
\begin{deluxetable}{llcccccl}
\tabletypesize{\footnotesize}
\tablenum{2}
\tablecolumns{10}
\tablecaption{Foreground Stars}
\label{tab:spec_fg}
\tablewidth{0pc}
\tablehead{\colhead{Catalog ID} & \colhead{Star ID} & \colhead{$\alpha_{J2000}$} & \colhead{$\delta_{J2000}$} & \colhead{$V$} & \colhead{$(B-V)$} & \colhead{Variability} & \colhead{Spectral Type}}

\startdata
  ---                  &  9492$\_$12$\_$5654   &  14:02:23.33  &  54:28:01.68 & ---   & ---  & $UB$  & M4 V        \\
  ---                  &  9490$\_$c2$\_$12822  &  14:02:39.23  &  54:19:16.83 & ---   & ---  & $UBVR$& K7 V        \\
  J140243.90+541727.85 &  B53                  &  14:02:43.90  &  54:17:27.85 & 21.30 & 0.58 &---    & F0-F2 III   \\
  J140247.62+541728.90 &  9490$\_$c2$\_$1281   &  14:02:47.62  &  54:17:28.90 & 19.87 & 0.43 &---    & Late-A      \\
  J140247.76+542833.56 &  9492$\_$09$\_$25840  &  14:02:47.76  &  54:28:33.56 & 19.91 & 0.57 & $U$   & F8 III/V    \\
  ---                  &  9492$\_$09$\_$27729  &  14:02:48.88  &  54:28:48.76 & ---   & ---  & $UBVR$& M2 V        \\
  J140250.09+542138.09 &  9490$\_$02$\_$2795   &  14:02:50.09  &  54:21:38.05 & 20.11 & 0.76 & $R$   & G8 V        \\
  J140301.20+541839.71 &  B65                  &  14:03:01.20  &  54:18:39.71 & 18.96 & 0.31 &---    & Late A III  \\
  J140310.60+541809.18 &  9490$\_$a2$\_$17826  &  14:03:10.60  &  54:18:09.18 & 22.45 & 0.47 &---    & A3 V (WD)   \\
  J140312.62+542056.72 &  9490$\_$01$\_$4552   &  14:03:12.62  &  54:20:56.72 & 19.56 & 0.34 &---    & F2 III      \\
  J140318.13+542400.97 &  9490$\_$03$\_$3062   &  14:03:18.13  &  54:24:00.97 & 19.04 & 0.69 &---    & G5 V        \\
  ---                  &  9490$\_$01$\_$4409   &  14:03:31.38  &  54:20:53.67 & ---   & ---  &---    & G4 V        \\
  J140335.25+542242.89 &  9490$\_$c1$\_$3304   &  14:03:35.25  &  54:22:42.89 & 22.10 & 1.60 & $U$   & M5 V        \\
  ---                  &  9490$\_$b2$\_$21     &  14:03:36.93  &  54:14:18.72 & ---   & ---  &---    & F8 V        \\
  J140342.55+541740.49 &  9490$\_$a1$\_$1828   &  14:03:42.55  &  54:17:40.49 & 19.03 & 0.79 & $BV$  & F8 V        \\
  ---                  &  9490$\_$c1$\_$1056   &  14:03:49.06  &  54:21:49.73 & ---   & ---  & $BVR$ & M2 V        \\
  ---                  &  9490$\_$c1$\_$3679   &  14:03:49.62  &  54:23:08.74 & ---   & ---  & $U$   & M0 V        \\
  ---                  &  9490$\_$b1$\_$11450  &  14:03:55.82  &  54:20:58.20 & ---   & ---  &---    & M4 V        \\
  J140356.08+542149.39 &  9490$\_$c1$\_$1033   &  14:03:56.08  &  54:21:49.32 & 19.08 & 0.90 & $B$   & G8 V        \\
\enddata
\tablecomments{Spectroscopic targets observed with Hectospec on the MMT.  Units of right ascension are hours, minutes, and seconds;  units of declination are degrees, minutes, and seconds.  Sources are sorted by increasing RA.  The \textit{HST}/ACS magnitudes, for recovered sources, are provided.}
\end{deluxetable}

%% file: table3.tex
\begin{deluxetable}{llccccccl}
\tabletypesize{\footnotesize}
\tablenum{3}
\tablecolumns{8}
\tablecaption{Remaining Targets}
\label{tab:LBTtargs}
\tablewidth{0pc}
\tablehead{\colhead{Catalog ID} & \colhead{Star ID} & \colhead{$\alpha_{J2000}$} & \colhead{$\delta_{J2000}$} & \colhead{$V$} & \colhead{$(B-V)$} & \colhead{Variability} & \colhead{LBT/MODS Field}}

\startdata
J140219.79+542315.29   &   9492$\_$13$\_$7415   &   14:02:19.79   &   54:23:15.29   &   20.26   &   0.16 & $VR$  &\\
J140219.85+542313.67   &   9492$\_$13$\_$7103   &   14:02:19.85   &   54:23:13.67   &   20.37   &   0.82 & $VR$  &\\
J140220.43+542313.06   &   9492$\_$13$\_$6986   &   14:02:20.43   &   54:23:13.06   &   20.48   &   0.31 & $BVR$ & Field 4 \\
J140221.34+542333.72   &   9492$\_$13$\_$11152  &   14:02:21.34   &   54:23:33.72   &   20.63   &   1.05 & $BVR$ & Field 4 \\
J140225.55+541917.18   &   9492$\_$14$\_$1816   &   14:02:25.55   &   54:19:17.18   &   20.85   &   0.15 &---    & Field 4 \\
J140226.24+541944.04   &   9492$\_$14$\_$6605   &   14:02:26.24   &   54:19:44.04   &   20.68   &   0.09 &---    & Field 4 \\
J140226.53+542335.74   &   9492$\_$13$\_$11533  &   14:02:26.53   &   54:23:35.74   &   20.23   &   1.41 & $UBVR$& Field 4 \\
J140226.64+541948.25   &   9492$\_$14$\_$7655   &   14:02:26.64   &   54:19:48.25   &   20.74   &   0.08 &---    &\\
J140226.75+541945.84   &   9492$\_$14$\_$7028   &   14:02:26.75   &   54:19:45.84   &   20.54   &   0.49 &---    &\\
J140226.97+541951.38   &   9492$\_$14$\_$8521   &   14:02:26.97   &   54:19:51.38   &   20.78   &   0.06 &---    &\\
J140227.03+541947.64   &   9492$\_$14$\_$7490   &   14:02:27.03   &   54:19:47.64   &   21.02   &   0.01 &---    &\\
J140227.68+542619.32   &   9492$\_$12$\_$961    &   14:02:27.68   &   54:26:19.32   &   20.94   &   0.20 & $UBVR$& Field 4 \\
J140231.47+542531.44   &   9492$\_$13$\_$23814  &   14:02:31.47   &   54:25:31.44   &   20.78   &   0.82 & $UBVR$&\\
J140232.51+542001.39   &   9492$\_$14$\_$11163  &   14:02:32.51   &   54:20:01.39   &   20.09   &   0.47 & $UBVR$&\\
J140234.99+542416.67   &   9492$\_$13$\_$17081  &   14:02:34.99   &   54:24:16.67   &   20.53   &   1.35 & $UBVR$& Field 4 \\
---                    &   9492$\_$14$\_$1739   &   14:02:39.24   &   54:19:16.77   &   ---     &   ---  & $UBVR$& \\
J140248.90+541840.10   &   9490$\_$c2$\_$10368  &   14:02:48.90   &   54:18:40.10   &   20.74   &   0.16 &---    &\\
J140249.10+541842.08   &   9490$\_$c2$\_$10564  &   14:02:49.10   &   54:18:42.08   &   20.24   &   0.13 &---    &\\
J140249.39+542359.14   &   9490$\_$b1$\_$16197  &   14:02:49.39   &   54:23:59.14   &   21.02   &   0.31 &---    & Field 3 \\
J140249.83+541924.17   &   9492$\_$09$\_$3069   &   14:02:49.83	  &   54:19:24.20   &   20.88   &   0.25 &---    & Field 1 \\ 
J140250.63+542346.86   &   9490$\_$a3$\_$15563  &   14:02:50.63   &   54:23:46.86   &   20.63   &   0.06 &---    & Field 3 \\
J140251.11+542114.51   &   9490$\_$03$\_$4663   &   14:02:51.11   &   54:21:14.51   &   20.60   &   0.49 & $U$   &\\
J140253.82+542319.14   &   9490$\_$a1$\_$5661   &   14:02:53.82   &   54:23:19.14   &   20.86   &  -0.04 &---    & Field 3 \\
J140254.07+541942.89   &   9490$\_$02$\_$867    &   14:02:54.07   &   54:19:42.89   &   20.73   &   0.04 & $U$   & Field 1 \\
J140256.45+541830.42   &   9490$\_$a2$-$SF10    &   14:02:56.45   &   54:18:30.42   &   20.40   &   0.00 &---    & Field 1 \\
J140257.81+541750.64   &   9490$\_$a3$\_$1058   &   14:02:57.81   &   54:17:50.64   &   20.85   &   0.42 &---    & Field 1 \\
J140258.34+541656.35   &   9490$\_$a2$\_$7168   &   14:02:58.34   &   54:16:56.35   &   20.88   &   0.72 &---    & Field 1 \\
J140258.67+542242.49   &   9490$\_$01$\_$7533   &   14:02:58.67   &   54:22:42.49   &   20.77   &   0.11 &---    &\\
                       &   9490$\_$a1$\_$6666   &   14:02:59.11   &   54:21:09.22   &   ---     &  ---   & $VR$  &\\
---                    &   9490$\_$a3$\_$21186  &   14:02:59.99   &   54:24:36.05   &   ---     &   ---  & $UBVR$& Field 3 \\
J140300.13+542156.52   &   9490$\_$02$\_$3047   &   14:03:00.13   &   54:21:56.52   &   20.47   &   0.12 &---    &\\
J140301.82+541514.87   &   9490$\_$a2$\_$39     &   14:03:01.82   &   54:15:14.87   &   20.86   &   0.51 &---    & Field 1 \\
J140302.00+542329.58   &   9492$\_$10$-$SF89    &   14:03:02.00   &   54:23:29.58   &   20.22   &   0.18 & $UBVR$& Field 3 \\
J140302.70+542308.88   &   9492$\_$10$-$SF162   &   14:03:02.70   &   54:23:08.88   &   21.38   &   0.92 &---    & Field 3 \\
J140306.60+541924.56   &   9490$\_$02$\_$584    &   14:03:06.60   &   54:19:24.56   &   20.68   &   0.14 &---    & Field 1 \\
J140306.84+542147.95   &   9490$\_$02$\_$2905   &   14:03:06.84   &   54:21:47.95   &   19.96   &  -0.06 &---    &\\
J140309.76+542330.52   &   9490$\_$a3$\_$12519  &   14:03:09.76   &   54:23:30.52   &   20.35   &   0.03 &---    & Field 3 \\
J140311.11+541623.59   &   9490$\_$a2$-$SF112   &   14:03:11.11   &   54:16:23.59   &   21.12   &   0.05 &---    & Field 1 \\
J140312.18+542006.40   &   9490$\_$02$\_$1378   &   14:03:12.18   &   54:20:06.40   &   20.42   &   0.01 &---    &\\
J140312.52+541752.55   &   9490$\_$a2$\_$15263  &   14:03:12.52   &   54:17:52.55   &   20.92   &   0.09 &---    &\\
J140312.62+541749.85   &   9490$\_$a2$\_$14783  &   14:03:12.62   &   54:17:49.85   &   20.70   &   0.20 & $V$   & Field 1 \\
J140312.81+541732.17   &   9490$\_$a2$\_$12011  &   14:03:12.81   &   54:17:32.17   &   20.84   &   0.13 &---    &\\
J140315.13+542151.19   &   9490$\_$01$\_$7652   &   14:03:15.13   &   54:21:51.19   &   20.62   &   0.13 &---    & Field 3 \\
J140315.66+541907.72   &   9492$\_$10$\_$5190   &   14:03:15.66   &   54:19:07.72   &   21.30   &  -0.01 &---    & Field 1 \\
J140316.03+541756.22   &   9490$\_$a2$\_$15969  &   14:03:16.03   &   54:17:56.22   &   20.28   &   0.23 & $U$   & Field 1 \\
J140317.37+542506.89   &   9490$\_$03$\_$6586   &   14:03:17.37   &   54:25:06.89   &   20.47   &   1.34 & $U$   & Field 3 \\
J140317.62+542332.53   &   9490$\_$03$-$SF55    &   14:03:17.62   &   54:23:32.53   &   21.27   &   0.06 &---    & Field 3 \\
J140317.99+541712.08   &   9490$\_$a2$\_$9284   &   14:03:17.99   &   54:17:12.02   &   21.29   &   0.00 &---    & Field 1 \\
---                    &   9490$\_$a3$\_$4912   &   14:03:19.25   &   54:21:49.12   &   ---     &   ---  & $UBVR$& Field 3 \\
J140320.92+541705.60   &   9490$\_$a2$\_$8459   &   14:03:20.92   &   54:17:05.60   &   20.53   &   0.38 & $U$   &\\
J140321.42+542344.09   &   9490$\_$03$\_$1988   &   14:03:21.42   &   54:23:44.09   &   20.46   &   1.26 &---    &\\
J140321.81+542346.00   &   9490$\_$03$\_$2120   &   14:03:21.81   &   54:23:46.00   &   20.31   &   0.11 &---    & Field 3 \\
J140323.24+542042.68   &   9490$\_$a2$\_$17072  &   14:03:23.24   &   54:20:42.68   &   20.82   &   0.30 &---    & Field 2 \\
J140324.77+541726.74   &   9490$\_$a1$\_$1294   &   14:03:24.77   &   54:17:26.74   &   20.90   &   0.15 &---    & Field 1 \\
J140325.40+541939.04   &   9490$\_$a1$\_$6386   &   14:03:25.40   &   54:19:39.04   &   20.59   &   1.40 & $BVR$ &\\
J140325.45+542522.94   &   9490$\_$03$\_$7146   &   14:03:25.45   &   54:25:22.94   &   20.66   &   0.02 & $V$   & Field 3 \\
J140325.57+542353.02   &   9490$\_$03$\_$2492   &   14:03:25.57   &   54:23:53.02   &   20.29   &   0.64 & $V$   & Field 3 \\
J140325.58+541957.65	 &   9490$\_$01$\_$1198   &   14:03:25.58	  &   54:19:57.60   &   20.58   &   0.10 & $VR$  & Field 2 \\
J140325.70+542514.88   &   9490$\_$03$\_$6836   &   14:03:25.70   &   54:25:14.88   &   20.27   &   0.04 & $R$   & Field 3 \\
J140325.90+542420.88   &   9490$\_$03$\_$4390   &   14:03:25.90   &   54:24:20.88   &   22.61   &   0.01 &---    & Field 3 \\
J140325.98+542422.82   &   9490$\_$03$\_$4520   &   14:03:25.98   &   54:24:22.82   &   20.02   &   0.41 &---    &\\
J140326.19+541939.47   &   9490$\_$c1$\_$1362   &   14:03:26.19   &   54:19:39.47   &   20.81   &   0.66 &---    & Field 2 \\
J140326.21+542047.87   &   9490$\_$01$\_$4031   &   14:03:26.21   &   54:20:47.87   &   20.67   &   0.01 & $U$   &\\
J140326.38+542411.30   &   9490$\_$02$\_$2637   &   14:03:26.38   &   54:24:11.30   &   20.69   &  -0.04 &---    & Field 3 \\
J140326.45+542038.36   &   9490$\_$01$\_$3379   &   14:03:26.45   &   54:20:38.36   &   20.65   &   0.07 &---    &\\
J140327.03+542346.36   &   9490$\_$03$\_$2142   &   14:03:27.03   &   54:23:46.36   &   21.08   &  -0.06 &---    &\\
J140327.10+542046.79   &   9490$\_$01$\_$3977   &   14:03:27.10   &   54:20:46.79   &   20.82   &   0.09 &---    & Field 2 \\
J140327.60+541846.80   &   9490$\_$a1$\_$4198   &   14:03:27.60   &   54:18:46.80   &   20.75   &   0.01 &---    &\\
J140327.67+542340.92   &   9490$\_$03$\_$1822   &   14:03:27.67   &   54:23:40.92   &   19.69   &   0.40 &---    &\\
J140327.71+541844.93   &   9490$\_$a1$\_$4062   &   14:03:27.71   &   54:18:44.93   &   20.89   &   0.05 &---    &\\
J140328.82+542458.03   &   9490$\_$03$\_$6315   &   14:03:28.82   &   54:24:58.03   &   20.15   &   0.84 & $U$   &\\
J140329.43+541809.97   &   9490$\_$a3$\_$7830   &   14:03:29.43   &   54:18:09.97   &   20.54   &   0.25 &---    &\\
J140329.51+541712.23   &   9490$\_$a1$\_$627    &   14:03:29.51   &   54:17:12.23   &   20.81   &   0.19 &---    &\\
J140330.07+541853.28   &   9490$\_$c1$\_$1485   &   14:03:30.07   &   54:18:53.28   &   20.52   &   0.30 & $U$   & Field 2 \\
J140330.32+541954.19   &   9490$\_$01$\_$991    &   14:03:30.32   &   54:19:54.19   &   20.80   &  -0.06 &---    &\\
J140330.61+542424.48   &   9490$\_$c2$\_$2831   &   14:03:30.61   &   54:24:24.48   &   20.45   &   0.77 & $UV$  & Field 3 \\
---                    &   9490$\_$b2$\_$14     &   14:03:30.91   &   54:14:10.48   &   ---     &   ---  &---    &\\
---                    &   9490$\_$b1$\_$14323  &   14:03:32.13   &   54:22:00.47   &   ---     &   ---  &---    & Field 2 \\
J140333.74+541854.61   &   9490$\_$a1$\_$4710   &   14:03:33.74   &   54:18:54.61   &   20.57   &   0.48 &---    & Field 2 \\
---                    &   9490$\_$b1$\_$3388   &   14:03:40.92   &   54:19:02.68   &   ---     &   ---  &---    & Field 2 \\
---                    &   9490$\_$03$\_$3779   &   14:03:41.57   &   54:19:08.47   &   ---     &   ---  & $UB$  &\\
J140342.85+542336.96   &   9490$\_$c1$\_$4050   &   14:03:42.85   &   54:23:36.96   &   20.65   &   0.20 & $UBV$ &\\
J140346.10+541959.52   &   9490$\_$b1$\_$8183   &   14:03:46.10   &   54:19:59.52   &   20.77   &   0.07 &---    & Field 2 \\
---                    &   9490$\_$c1$\_$1535   &   14:03:46.17   &   54:16:15.68   &   ---     &   ---  & $UBVR$&\\
---                    &   9490$\_$c1$\_$3069   &   14:03:47.38   &   54:22:29.50   &   ---     &   ---  & $UBVR$&\\
---                    &   9490$\_$c1$\_$3679   &   14:03:49.63   &   54:23:8.74    &   ---     &   ---  & $BV$  &\\
J140351.93+542152.78   &   9490$\_$c1$\_$1292   &   14:03:51.93   &   54:21:52.78   &   20.33   &  -0.02 &---    &\\
J140355.88+542229.46   &   9490$\_$c1$\_$3067   &   14:03:55.88   &   54:22:29.46   &   20.30   &   0.11 &---    &\\
\enddata

\end{deluxetable}